\documentclass[11pt]{article}
\usepackage{amssymb,amsmath,amsfonts}
\usepackage{graphicx}
\usepackage{graphics}
\usepackage{eepic,epsfig}

\@addtoreset{equation}{section}

\makeatother

\begin{document}

\title{Vacuum densities and the Casimir forces for branes\\
orthogonal to the AdS boundary}
\author{S. Bellucci$^{1}$\thanks{%
E-mail: bellucci@lnf.infn.it },\, A. A. Saharian$^{2}$\thanks{%
E-mail: saharian@ysu.am},\, V. Kh. Kotanjyan$^{2,3}$\thanks{%
E-mail: vatokoto@gmail.com} \\
\\
\textit{$^1$ INFN, Laboratori Nazionali di Frascati,}\\
\textit{Via Enrico Fermi 54, 00044 Frascati, Italy} \vspace{0.3cm}\\
\textit{$^2$Department of Physics, Yerevan State University,}\\
\textit{1 Alex Manoogian Street, 0025 Yerevan, Armenia} \vspace{0.3cm}\\
\textit{$^3$Institute of Applied Problems of Physics NAS RA,}\\
\textit{25 Hrachya Nersissyan Street, 0014 Yerevan, Armenia}}
\maketitle

\begin{abstract}
For a massive scalar field with general curvature coupling we evaluate the
Wightman function in the geometry of two parallel branes perpendicular to
the AdS boundary. On the separate branes, the field operator is constrained
by Robin boundary conditions, in general, with different coefficients. In
the region between the branes their contribution to the Wightman function is
explicitly separated. By using this decomposition, the brane-induced effects
on the vacuum expectation values (VEVs) for the field squared and
energy-momentum tensor are investigated. The behavior of those expectation
values is studied in various asymptotic regions of the parameters. The
vacuum energy-momentum tensor in addition to the diagonal components has a
nonzero off-diagonal stress. Depending on the boundary conditions and also
on the distance from the branes, the vacuum energy density can be either
positive or negative. The Casimir forces acting on the branes have two
components. The first one corresponds to the standard normal force and the
second one is parallel to the branes and presents the vacuum shear force.
Unlike to the problem of parallel plates in the Minkowski bulk, the normal
Casimir forces acting on separate branes differ if the boundary conditions
on the branes are different. They can be either repulsive or attractive. In
a similar way, depending on the coefficients in the boundary conditions, the
shear force is directed toward or from the AdS boundary. The separate
components may also change their signs as functions of the interbrane
separation. At large proper separations between the branes, compared to the
AdS curvature radius, both the components of the Casimir forces exhibit a
power-law decay. For a massive scalar field this behavior is in contrast to
that for the Minkowski bulk, where the decrease is exponential.
\end{abstract}

Keywords: Casimir effect, AdS spacetime, branes

\bigskip

\section{Introduction}

\label{sec:Int}

Motivated by fundamental questions and applications in condensed matter
physics, cosmology and in high-energy physics, the Casimir effect (for
reviews see \cite{Most97}) remains to be an active field of research in
quantum field theory. The effect is an interesting manifestation of quantum
fluctuations of fields influenced by the presence of boundaries or by
nontrivial spatial topology. Depending on the model under consideration the
physical nature of the boundaries can be different. Examples are the
interfaces of macroscopic bodies in quantum electrodynamics, boundaries
separating different phases of the system, horizons in gravitational
physics, branes in string theories and in cosmological models of braneworld
type, etc. The boundary and periodicity conditions imposed on the operator
of a quantum field modify the spectrum of fluctuations and result in the
shift of the expectation values of physical quantities such as
energy-momentum tensor or current densities for charged fields.

In addition to the boundary or periodicity conditions imposed on the field,
the properties of quantum fluctuations are sensitive to the presence of
background classical fields. Those fields reduce the symmetry in respective
problems and exact results for physical characteristics in the Casimir
effect are obtained for highly symmetric bulk and boundary geometries only.
In the present paper we consider the influence of the background
gravitational field on the properties of the scalar vacuum in the geometry
of two parallel branes in background of anti-de Sitter (AdS) spacetime. That
geometry is the maximally symmetric solution of the Einstein field equations
with a negative cosmological constant as the only source of the
gravitational field. As it will be shown below, this high symmetry allows to
obtain closed analytic expressions for the expectation values characterizing
the local properties of the vacuum state. In addition to the high symmetry,
our choice of AdS spacetime as the background geometry is motivated by its
important role in two exciting developments of theoretical physics during
the last decade, namely, AdS/conformal field theory (CFT) correspondence and
braneworld scenarios with large extra dimensions. The AdS/CFT correspondence
(for reviews see, e.g., \cite{Ahar00}) establishes duality between two
different theories: supergravity or string theory on asymptotically AdS bulk
from one side and conformal field theory on AdS boundary from another one.
Those theories live in different numbers of spacetime dimensions and the
correspondence is an example of holographic duality. It provides an
important possibility to investigate strong coupling non-perturbative
effects in one theory by mapping them to weak coupling region of dual theory
and has been applied in different physical settings including the variety of
condensed matter systems. The braneworld paradigm \cite{Maar10} naturally
arises in the context of supergravity and string theories and presents an
alternative to Kaluza-Klein compactification of extra dimensions. The models
formulated on AdS bulk provide a geometrical solution for the hierarchy
problem between the electroweak and gravitational energy scales and also new
perspectives and different interpretations for various problems in particle
physics and cosmology.

In the Randall-Sundrum-type realizations of the braneworld models \cite%
{Rand99} the branes are parallel to the AdS boundary. Motivated by the
radion stabilization and the generation of cosmological constant on branes,
the Casimir effect in that setup has been widely investigated in the
literature for scalar \cite{Fabi00,Flac01}, fermionic \cite{Flac01f} and
vector \cite{Garr03} fields. In the main part of the papers, as a physical
characteristic of the vacuum, global quantities, such as the Casimir energy
or the effective potential, are investigated by using various regularization
schemes. Local observables carry more detailed information about the
properties of the vacuum state. In particular, being a source of gravity in
semiclassical Einstein equations, the vacuum expectation value (VEV) of the
energy-momentum tensor is of special importance. It is investigated in \cite%
{Saha03,Saha05, Saha07} for scalar, fermionic and electromagnetic fields.
The combined effects of a brane and topological defect of a cosmic string
type on the local characteristics of the fermionic vacuum in AdS spacetime
have been recently considered in \cite{Bell22}. For charged fields, another
important local characteristic of the vacuum state, bilinear in the field,
is the VEV of the current density. It has been studied in \cite{Bell17} for
scalar and fermionic fields in the geometry of branes parallel to the
boundary of locally AdS spacetime with a part of spatial dimensions
compactified to a torus.

Motivated by an increase of interest to conformal field theories in the
presence of boundaries (see, for example, references given in \cite{Cuom21}%
), in recent studies the AdS/CFT correspondence is extended to the problems
where boundaries are present in the conformal field theory side. In the
corresponding setup the boundary CFT is dual to a theory in AdS bulk with
additional boundaries intersecting the AdS boundary at the locations of
boundaries in CFT (AdS/BCFT correspondence) \cite{Taka11}. Problems with
surfaces in the AdS bulk crossing the AdS boundary have been considered in
recent studies of entanglement entropy in the context of AdS/CFT
correspondence \cite{Ryu06} (for reviews see \cite{Nish09,Chen22}). A
geometric classical procedure is suggested for evaluation of the
entanglement entropy of quantum systems living on the AdS boundary. In
accordance of that procedure, the entanglement entropy for a bounded region
in CFT is expressed in terms of the area of the minimal surface in the AdS
bulk anchored at the boundary of that region.

In the papers cited above, the physical characteristics in the Casimir
effect with branes serving as constraining boundaries, have been considered
in the context of Randall-Sundrum-type models with branes parallel to the
AdS boundary. Motivated by recent developments for physical models on AdS
bulk with boundaries crossing the spacetime boundary and continuing the
investigation started in \cite{Beze15}, in the present paper we consider a
problem with two branes orthogonal to the AdS boundary for a massive scalar
field with general curvature coupling parameter. Though this problem is less
symmetric than the setups with branes parallel to the AdS boundary, as it
will be seen below, it is still exactly solvable.

The organization of the paper is as follows. In the next section we fix the
problem setup and present the complete set of mode functions for a scalar
field in the region between the branes. By using those functions, the
positive frequency Wightman function is evaluated in Section \ref{sec:WF}.
The brane-induced contribution is explicitly separated. Taking the
coincidence limit of the arguments in that contribution, the mean field
squared is investigated in Section \ref{sec:phi2}. The behavior of the VEV
in various asymptotic regions for the values of the parameters is discussed.
Similar investigations for the VEV of the energy-momentum tensor are
presented in Section \ref{sec:EMT}. The Casimir forces acting on the branes
are discussed in Section \ref{sec:Forces}. It is shown that for Robin
boundary conditions, in addition to the normal component, those forces have
a nonzero component parallel to the branes (shear force). The nature of the
forces is studied depending on the boundary conditions. The main results are
summarized in Section \ref{sec:Conc}. In Appendix \ref{sec:Ap}, by using a
variant of the generalized Abel-Plana formula we provide an integral
representation for the series in the mode-sum over the eigenvalues of the
quantum number describing the degree of freedom along the direction normal
to the branes.

\section{Problem setup and the field modes}

\label{sec:mf}

We consider a scalar field $\varphi (x)$ on the background of a $(D+1)$%
-dimensional AdS spacetime with the curvature radius $\alpha $. In Poincar%
\'{e} coordinates the corresponding line element is given by
\begin{equation}
ds^{2}=g_{ik}dx^{i}dx^{k}=e^{-2y/\alpha }\left[ dt^{2}-\left( dx^{1}\right)
^{2}-d\mathbf{x}^{2}\right] -dy^{2},  \label{ds}
\end{equation}%
where the coordinates $\mathbf{x}=(x^{2},\ldots ,x^{D-1})$ are separated for
the future convenience. In addition to the coordinate $y$, $-\infty
<y<+\infty $, we will also use the coordinate $z$, defined as $z=\alpha
e^{y/a}$, $0<z<\infty $, in terms of which the line element is written in a
manifestly conformally flat form
\begin{equation}
ds^{2}=\left( \frac{\alpha }{z}\right) ^{2}\left[ dt^{2}-\left(
dx^{1}\right) ^{2}-d\mathbf{x}^{2}-dz^{2}\right] .  \label{dsc}
\end{equation}%
The AdS boundary and horizon are presented by the hypersurfaces $z=0$ and $%
z=\infty $, respectively. The Ricci scalar and the cosmological constant are
expressed in terms of the AdS\ curvature radius by the relations $%
R=-D(D+1)/\alpha ^{2}$ and $\Lambda =-D(D-1)/(2\alpha ^{2})$.

\begin{figure}[tbph]
\begin{center}
\epsfig{figure=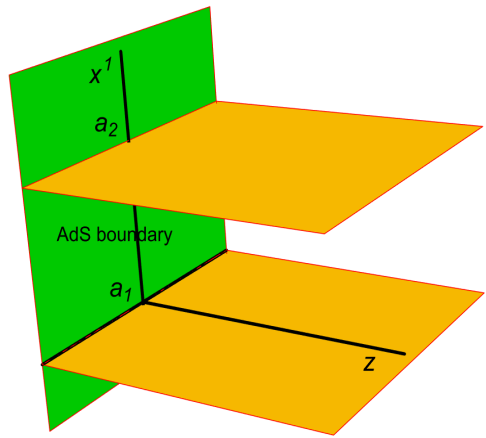,width=7cm,height=7.5cm}
\end{center}
\caption{The geometry of the problem with branes intersecting the AdS
boundary.}
\label{fig1}
\end{figure}

The operator of the scalar field with the curvature coupling constant $\xi $
obeys the equation
\begin{equation}
\left( g^{ik}\nabla _{i}\nabla _{k}+m^{2}+\xi R\right) \varphi (x)=0.
\label{feq}
\end{equation}%
The most popular special cases correspond to minimally and conformally
coupled fields with $\xi =0$ and $\xi =\xi _{D}=(D-1)/(4D)$, respectively.
We are interested in the effects of two branes located at $x^{1}=a_{1}$ and $%
x^{1}=a_{2}$ on the local properties of vacuum state for the field $\varphi
(x)$ (see Figure \ref{fig1}). It is assumed that on the brane at $%
x^{1}=a_{j} $, $j=1,2$, the field obeys Robin boundary condition
\begin{equation}
(A_{j}+B_{j}n_{j}^{i}\nabla _{i})\varphi (x)=0,  \label{bc1}
\end{equation}%
where $n_{j}^{i}$ is the normal to the brane. The discussion in what follows
will be mainly focused on the VEVs in the region between the branes, $%
a_{1}\leq x^{1}\leq a_{2}$, with $n_{j}^{i}=(-1)^{j-1}\delta
_{1}^{i}z/\alpha $. We will consider the special case with $%
B_{j}/A_{j}=\alpha \beta _{j}/z$, where $\beta _{j}$, $j=1,2$, are
constants. With this choice, the boundary condition (\ref{bc1}) in the
region between the branes is written as%
\begin{equation}
(1+(-1)^{j-1}\beta _{j}\partial _{1})\varphi (x)=0.  \label{bc2}
\end{equation}%
Note that for a given $z$, the physical coordinate that measures the proper
distance from the branes is given by $x_{(p)}^{1}=\alpha x^{1}/z$ and the
condition (\ref{bc1}) is presented as $(1+\beta _{j}n_{j}^{1}\partial
_{x_{(p)}^{1}})\varphi (x)=0$. This means that the coefficient in the Robin
boundary condition written in terms of the coordinate $x_{(p)}^{1}$ is
constant. The results for Dirichlet and Neumann boundary conditions are
obtained in the special cases $\beta _{j}=0$ and $\beta _{j}=\infty $,
respectively.

Our use of the term "brane" for the boundaries is, in some sense,
conditional. Fundamental branes in string theory or phenomenological branes
in braneworld scenarios are among the possible physical realizations of the
boundary conditions (\ref{bc1}). For example, in Randall-Sundrum type models
they follow from the $Z_{2}$-symmetry with respect to the branes and the
corresponding Robin coefficients are expressed in terms of constants in the
brane mass terms of the part of the action located on the branes (see \cite%
{Flac01,Saha05}). The Robin conditions also arise on boundaries separating
spatial regions with different geometries (this type of setup on the AdS
bulk has been considered in \cite{Saha07} to model the finite thickness of
branes). In this case the Robin coefficients are expressed in terms of
geometric characteristics of the contacting regions. The Robin type boundary
conditions were used to model the finite penetration of the field into the
boundary with the penetration length determined by the coefficient in the
boundary condition.

The properties of the vacuum state in the problem under consideration are
encoded in two-point functions. Those functions are presented in the form of
sums over complete set of the field modes obeying the boundary conditions.
Those modes for a scalar field in AdS spacetime when the branes are absent
are well known in the literature. We will denote by $\lambda $ the quantum
number corresponding to the coordinate $x^{1}$. For the problem at hand the
corresponding solutions are found by combining the factors $e^{i\lambda
x^{1}}$ and $e^{-i\lambda x^{1}}$ with the relative coefficient that will be
determined from the boundary conditions. Denoting by $\sigma $ the set of
quantum numbers specifying the modes, in the region between the branes the
mode functions are written in the form%
\begin{equation}
\varphi _{\sigma }(x)=C_{\sigma }z^{D/2}J_{\nu }(\gamma z)\cos [\lambda
|x^{1}-a_{j}|+\alpha _{j}(\lambda )]e^{i\mathbf{k}\mathbf{x}-iEt},
\label{mf}
\end{equation}%
where $J_{\nu }(u)$ is the Bessel function, $\mathbf{k}=(k^{2},\dots
,k^{D-1})$, $-\infty <k^{l}<+\infty $, $0\leq \lambda <+\infty $, and the
energy is expressed as
\begin{equation}
E=\sqrt{\lambda ^{2}+\mathbf{k}^{2}+\gamma ^{2}}.  \label{E}
\end{equation}%
In (\ref{mf}) and in what follows we use the notation
\begin{equation}
\nu =\sqrt{\frac{D^{2}}{4}-D(D+1)\xi +m^{2}\alpha ^{2}},  \label{nuJ}
\end{equation}%
assuming that $\nu \geq 0$. This condition is dictated by the stability of
the vacuum state \cite{Brey82}. With the mode functions (\ref{mf}) the set
of quantum numbers is specified as $\sigma =(\lambda ,\mathbf{k},\gamma )$.

From the boundary condition at $x^{1}=a_{j}$ it follows that%
\begin{equation}
e^{2i\alpha _{j}(\lambda )}=\frac{i\lambda \beta _{j}-1}{i\lambda \beta
_{j}+1}.  \label{alfj}
\end{equation}%
We have $\alpha _{j}(\lambda )=\pi /2$ and $\alpha _{j}(\lambda )=0$ for
Dirichlet and Neumann conditions, respectively. The boundary condition on
the second brane gives the equation that determines the eigenvalues for the
quantum number $\lambda $:%
\begin{equation}
\left( b_{1}+b_{2}\right) u\cos u+\left( u^{2}b_{1}b_{2}-1\right) \sin u=0,
\label{meq}
\end{equation}%
where $u=\lambda a$ and $b_{j}=\beta _{j}/a$. This eigenvalue equation is
the same as the corresponding equation for two parallel plates in the
Minkowski bulk, considered in \cite{Rome02}. As it has been discussed in
\cite{Rome02}, depending on the values of $b_{j}$, the equation (\ref{meq})
may have single or two purely imaginary roots with respect to $u$. In the
presence of those roots and for the part of the modes with $\mathbf{k}%
^{2}+\gamma ^{2}<|u|^{2}/a^{2}$ the energy becomes imaginary which signals
about the instability of the vacuum state. Note that here the situation is
different from that in the corresponding problem on the Minkowski bulk. In
the latter problem the mass enters in the expression for the energy, and for
imaginary modes with $|u|/a<m$ the energy is positive for all the modes and
the vacuum is stable. To have a stable vacuum state, we will assume the
values of the parameters $b_{1}$ and $b_{2}$ for which all the roots of the
equation (\ref{meq}) are real. Those values belong to the region in the
plane $(b_{1},b_{2})$ given by $\{b_{1}+b_{2}\geq 1,b_{1}b_{2}\leq 0\}\cup
\{b_{1,2}\leq 0\}$ (see \cite{Rome02}). We will denote by $u=u_{n}$, $%
n=1,2,\ldots $, the positive roots of the equation (\ref{meq}). For the
eigenvalues of the quantum number $\lambda $ one has $\lambda =\lambda
_{n}=u_{n}/a$. For Dirichlet and Neumann boundary conditions the eigenvalue
equation is reduced to $\sin u=0$ with the modes $\lambda _{n}=\pi n/a$,
where $n=1,2,\ldots $, and $n=0,1,2,\ldots $, for the first and second
cases, respectively. Note the presence of an additional zero mode for
Neumann condition. For Dirichlet condition on one brane and the Neumann one
on another from (\ref{meq}) we get $\cos u=0$ and $\lambda _{n}=\pi
(n-1/2)/a $, $n=1,2,\ldots $.

The constant $C_{\sigma }$ in (\ref{mf}) is determined from the
normalization condition%
\begin{equation}
\int d^{D}x\sqrt{|g|}g^{00}\varphi _{\sigma }(x)\varphi _{\sigma ^{\prime
}}^{\ast }(x)=\frac{\delta _{nn^{\prime }}}{2E}\delta (\mathbf{k}-\mathbf{k}%
^{\prime })\delta (\gamma -\gamma ^{\prime }).  \label{norm}
\end{equation}%
For the mode functions (\ref{mf}) this gives%
\begin{equation}
\left\vert C_{\sigma }\right\vert ^{2}=\frac{\left( 2\pi \right)
^{2-D}\gamma }{\alpha ^{D-1}aEN_{n}},  \label{C2}
\end{equation}%
with the notation%
\begin{equation}
N_{n}=1+\frac{\sin u_{n}}{u_{n}}\cos \left[ u_{n}+2\alpha _{j}(\lambda _{n})%
\right] .  \label{Nn}
\end{equation}%
Note that $\cos \left[ u_{n}+2\alpha _{1}(\lambda _{n})\right] =\cos \left[
u_{n}+2\alpha _{2}(\lambda _{n})\right] $. Having fixed the complete set of
modes we pass to the evaluation of the Wightman function.

\section{Wightman function}

\label{sec:WF}

As a two-point function we consider the positive frequency Wightman function
defined as the VEV $W(x,x^{\prime })=\langle 0|\varphi (x)\varphi (x^{\prime
})|0\rangle $. Expanding the operators $\varphi (x)$ and $\varphi (x^{\prime
})$ in terms of the complete set $\{\varphi _{\sigma }(x),\varphi _{\sigma
}^{\ast }(x)\}$ and using the definition of the vacuum state, it is written
in the form of the following sum over the modes:
\begin{equation}
W(x,x^{\prime })=\int_{0}^{\infty }d\gamma \int d\mathbf{k}%
\sum_{n=1}^{\infty }\varphi _{\sigma }(x)\varphi _{\sigma }^{\ast
}(x^{\prime }).  \label{Wf1}
\end{equation}%
The problem is homogeneous in the subspace $(t,\mathbf{x})$ and we expect
that the dependence on the arguments in that subspace will enter in the form
of the differences $\Delta t=t-t^{\prime }$ and $\Delta \mathbf{x=x}-\mathbf{%
x}^{\prime }$. Substituting the functions (\ref{mf}) and the expression (\ref%
{C2}) for the normalization coefficient, the Wightman function is expressed
as
\begin{equation}
W(x,x^{\prime })=\frac{2(zz^{\prime })^{D/2}}{(2\pi \alpha )^{D-1}a}\int d%
\mathbf{k\,}e^{i\mathbf{k}\Delta \mathbf{x}}\int_{0}^{\infty }d\gamma
\,\gamma J_{\nu }(\gamma z)J_{\nu }(\gamma z^{\prime })S(b,\Delta
t,x^{1},x^{\prime 1}),  \label{Wf2}
\end{equation}%
where $b=\sqrt{\gamma ^{2}+\mathbf{k}^{2}}$, and
\begin{equation}
S(b,\Delta t,x^{1},x^{\prime 1})=\pi \sum_{n=1}^{\infty }\frac{e^{-i\sqrt{%
\lambda _{n}^{2}+b^{2}}\Delta t}}{\sqrt{\lambda _{n}^{2}+b^{2}}N_{n}}\cos {%
[\lambda _{n}|x}^{1}{-a_{j}|+\alpha _{j}(\lambda _{n})]}\cos {[\lambda _{n}{%
|x}}^{\prime }{^{1}{-a_{j}|}+\alpha _{j}(\lambda _{n})]}.  \label{S1}
\end{equation}%
An equivalent representation for the series over $n$ is obtained by using
the definition (\ref{alfj}) for the function ${\alpha _{j}(\lambda )}$:
\begin{equation}
S(b,\Delta t,x^{1},x^{\prime 1})=\frac{\pi }{4}\sum_{n=1}^{\infty }\frac{%
e^{-i\sqrt{\lambda _{n}^{2}+b^{2}}\Delta t}}{\sqrt{\lambda _{n}^{2}+b^{2}}%
N_{n}}\left[ 2\cos \left( {\lambda _{n}{\Delta x}^{1}}\right) {+}\sum_{l=\pm
1}\left( {e}^{i{\lambda _{n}|{{x}^{1}{+x}}^{\prime }{^{1}-2{a_{j}}}|}}\frac{%
i\lambda \beta _{j}-1}{i\lambda \beta _{j}+1}\right) ^{l}\right] .
\label{S2}
\end{equation}%
For boundary conditions different from Dirichlet or Neumann ones on both the
branes, the eigenvalues ${\lambda _{n}}$ are given implicitly, as roots of (%
\ref{meq}), and the representation (\ref{Wf2}) with (\ref{S1}) or (\ref{S2})
is not convenient for the investigation of the local VEVs in the coincidence
limit.

In order to get around this inconvenience and also to separate explicitly
the divergence in the coincidence limit, the integral representation (\ref%
{S4}) for the function (\ref{S2}) is obtained in Appendix \ref{sec:Ap} by
using the generalized Abel-Plana formula from \cite{SahaRev}. Substituting (%
\ref{S4}) in (\ref{Wf2}), the Wightman function is decomposed as%
\begin{eqnarray}
W(x,x^{\prime }) &=&W_{j}(x,x^{\prime })+\frac{(zz^{\prime })^{\frac{D}{2}}}{%
(2\pi \alpha )^{D-1}}\int d\mathbf{k\,}e^{i\mathbf{k}\Delta \mathbf{x}%
}\int_{0}^{\infty }d\gamma \,\gamma J_{\nu }(\gamma z)J_{\nu }(\gamma
z^{\prime })  \notag \\
&&\times \int_{b}^{\infty }d\lambda \frac{\cosh (\sqrt{\lambda ^{2}-b^{2}}%
\Delta t)}{\sqrt{\lambda ^{2}-b^{2}}}\frac{2\cosh \left( {\lambda {x}_{-}^{1}%
}\right) {+}\sum_{l=\pm 1}\left[ {e}^{{|{{x}_{+}^{1}-2{a_{j}}}|\lambda }%
}c_{j}(\lambda )\right] ^{l}}{c_{1}(\lambda )c_{2}(\lambda )e^{2a\lambda }-1}%
,  \label{Wf3}
\end{eqnarray}%
where and in what follows $x_{\pm }^{1}=x^{1}\pm x^{\prime 1}$ and%
\begin{equation}
c_{j}(\lambda )=\frac{\beta _{j}\lambda -1}{\beta _{j}\lambda +1}.
\label{cj}
\end{equation}%
Note that $c_{j}(\lambda )=\tilde{c}_{j}(\lambda a)$, with the functions $%
\tilde{c}_{j}(u)$ defined in Appendix \ref{sec:Ap} after formula (\ref{APf}%
). In (\ref{Wf3}) we have defined the two-point function
\begin{eqnarray}
W_{j}(x,x^{\prime }) &=&W_{0}(x,x^{\prime })+\frac{(zz^{\prime })^{\frac{D}{2%
}}}{(2\pi \alpha )^{D-1}}\int d\mathbf{k\,}e^{i\mathbf{k}\Delta \mathbf{x}%
}\int_{0}^{\infty }d\gamma \,\gamma J_{\nu }(\gamma z)J_{\nu }(\gamma
z^{\prime })  \notag \\
&&\times \int_{b}^{\infty }d\lambda \frac{\cosh (\sqrt{\lambda ^{2}-b^{2}}%
\Delta t)}{\sqrt{\lambda ^{2}-b^{2}}}\frac{e^{-{|{{x}_{+}^{1}-2{a_{j}}}%
|\lambda }}}{c_{j}(\lambda )}.  \label{Wj}
\end{eqnarray}%
Here, the part $W_{0}(x,x^{\prime })$ comes from the term $S_{0}(b,\Delta t,{%
{x}_{-}^{1}})$ in (\ref{S3}) and is given by
\begin{equation}
W_{0}(x,x^{\prime })=\frac{(zz^{\prime })^{\frac{D}{2}}}{2(2\pi \alpha
)^{D-1}}\int d\mathbf{K\,}e^{i\mathbf{K}\Delta \mathbf{X}}\int_{0}^{\infty
}d\gamma \,\gamma J_{\nu }(\gamma z)J_{\nu }(\gamma z^{\prime })\frac{e^{-i%
\sqrt{\gamma ^{2}+\mathbf{K}^{2}}\Delta t}}{\sqrt{\gamma ^{2}+\mathbf{K}^{2}}%
},  \label{W0}
\end{equation}%
with $\mathbf{X}=(x^{1},\mathbf{x})$, $\mathbf{K}=(k^{1},k^{2},\ldots
,k^{D-1})$. The integration over the angular coordinates of the vector $%
\mathbf{k}$\ in (\ref{Wf3}) and (\ref{Wj}) (and in a similar way for (\ref%
{W0})) can be done by using the formula%
\begin{equation}
\int d\mathbf{k\,}e^{i\mathbf{k}\Delta \mathbf{x}}g(k)=\frac{\left( 2\pi
\right) ^{\frac{D}{2}-1}}{|\Delta \mathbf{x}|^{\frac{D}{2}-2}}%
\int_{0}^{\infty }dk\,k^{\frac{D}{2}-1}J_{\frac{D}{2}-2}(k|\Delta \mathbf{x}%
|)g(k),  \label{IntAng}
\end{equation}%
for a given function $g(k)$, where $k=|\mathbf{k}|$.

The separate terms in (\ref{Wf3}) and (\ref{Wj}) have clear physical
interpretation. The function $W_{0}(x,x^{\prime })$ is the Wightman function
in AdS spacetime in the absence of the branes. Its expression in terms of
the hypergeometric function is well known from the literature (see below).
As it has been mentioned in Appendix \ref{sec:Ap}, the last term in (\ref%
{Wf3}) vanishes in the limit $(-1)^{j^{\prime }}a_{j^{\prime }}\rightarrow
+\infty $, where $j^{\prime }=1$ for $j=2$ and $j^{\prime }=2$ for $j=1$.
Hence, the function $W_{j}(x,x^{\prime })$ corresponds to the Wightman
function in the problem with a single brane at $x^{1}=a_{j}$. It has been
obtained in \cite{Beze15}. The last term in (\ref{Wf3}) is interpreted as a
contribution induced by the second brane at $x^{1}=a_{j^{\prime }}$ when we
add it to the geometry with a brane at $x^{1}=a_{j}$. The representation of
the Wightman function with combined contributions from the branes is
obtained from (\ref{S5}):%
\begin{eqnarray}
W(x,x^{\prime }) &=&W_{0}(x,x^{\prime })+\frac{(zz^{\prime })^{\frac{D}{2}}}{%
(2\pi \alpha )^{D-1}}\int d\mathbf{k\,}e^{i\mathbf{k}\Delta \mathbf{x}%
}\int_{0}^{\infty }d\gamma \,\gamma J_{\nu }(\gamma z)J_{\nu }(\gamma
z^{\prime })  \notag \\
&&\times \int_{b}^{\infty }d\lambda \frac{\cosh (\sqrt{\lambda ^{2}-b^{2}}%
\Delta t)}{\sqrt{\lambda ^{2}-b^{2}}}\frac{2\cosh \left( {\lambda {x}_{-}^{1}%
}\right) {+}\sum_{j=1,2}{e}^{{|{{x}_{+}^{1}-2{a_{j}}}|\lambda }%
}c_{j}(\lambda )}{c_{1}(\lambda )c_{2}(\lambda )e^{2a\lambda }-1}.
\label{Wf4}
\end{eqnarray}%
For the angular part of the integral over $\mathbf{k}$\ we can use the
relation (\ref{IntAng}).

In the representations (\ref{Wf3}) and (\ref{Wf4}) the explicit knowledge of
the eigenvalues $\lambda _{n}$ is not required and they are well adapted for
the investigation of local VEVs. Those representations give the Wightman
function in the region between the branes. In the regions $x^{1}<a_{1}$ and $%
x^{1}>a_{2}$ the Wightman functions coincide with that for the problem with
a single brane and they are given by (\ref{Wj}) with $j=1$ and $j=2$ in the
first and second regions respectively.

For the special cases of Dirichlet and Neumann boundary conditions we have $%
c_{j}(\lambda )=-\delta _{\mathrm{J}}$, where $\mathrm{J}=\mathrm{D},\mathrm{%
N}$ correspond to Dirichlet and Neuamann boundary conditions with $\delta _{%
\mathrm{D}}=1$, $\delta _{\mathrm{N}}=-1$. The part with the exponential
function is reduced to $1/(e^{2a\lambda }-1)$. Presenting this function as
the series $\sum_{n=1}^{\infty }e^{-2na\lambda }$, the integral over $%
\lambda $ is expressed in terms of the modified Bessel function $K_{0}(u)$
(see \cite{Grad07}). Next, we use the result (\ref{IntAng}) for the integral
over the angular coordinates of $k$. The integral over $k$ is expressed
through the associated Legendre function $Q_{\beta }^{\mu }(x)$ and the
final expression reads%
\begin{equation}
W(x,x^{\prime })=W_{0}(x,x^{\prime })+\frac{\alpha ^{1-D}}{2^{\frac{D}{2}%
+\nu +1}\pi ^{\frac{D}{2}}}\sum_{n=1}^{\infty }\left[ \sum_{l=\pm 1}f_{\nu
}(u_{l,n}^{(-)}){-}\delta _{\mathrm{J}}\sum_{j=1,2}f_{\nu }(u_{j,n}^{(+)})%
\right] ,  \label{WDN}
\end{equation}%
with the notations%
\begin{eqnarray}
u_{l,n}^{(-)} &=&1+\frac{\left( 2nla-{{x}_{-}^{1}}\right) ^{2}+|\Delta
\mathbf{x}|^{2}+\Delta z^{2}-\Delta t^{2}}{2zz^{\prime }},  \notag \\
u_{j,n}^{(+)} &=&1+\frac{\left( 2na-{|{{x}_{+}^{1}-2{a_{j}}}|}\right)
^{2}+|\Delta \mathbf{x}|^{2}+\Delta z^{2}-\Delta t^{2}}{2zz^{\prime }},
\label{ulj}
\end{eqnarray}%
and $\Delta z=z-z^{\prime }$. In (\ref{WDN}) we have defined the function%
\begin{eqnarray}
f_{\nu }(u) &=&\frac{2^{\nu +\frac{1}{2}}}{\sqrt{\pi }}e^{-\frac{D-1}{2}\pi
i}\frac{Q_{\nu -\frac{1}{2}}^{\frac{D-1}{2}}(u)}{\left( u^{2}-1\right) ^{%
\frac{D-1}{4}}}  \notag \\
&=&\frac{\Gamma \left( \nu +\frac{D}{2}\right) }{\Gamma \left( \nu +1\right)
u^{\nu +\frac{D}{2}}}\,F\left( \frac{2+2\nu +D}{4},\frac{2\nu +D}{4};\nu +1;%
\frac{1}{u^{2}}\right) ,  \label{fnu}
\end{eqnarray}%
with $F(a,b;c;x)\equiv \,_{2}F_{1}(a,b;c;x)$ being the hypergeometric
function.

The Wightman function for a scalar field in the brane-free AdS spacetime is
expressed in terms of the function $f_{\nu }(u)$ as%
\begin{equation}
W_{0}(x,x^{\prime })=\frac{\alpha ^{1-D}f_{\nu }(u_{0,0}^{(-)})}{2^{\frac{D}{%
2}+\nu +1}\pi ^{\frac{D}{2}}},  \label{W01}
\end{equation}%
and the formula (\ref{WDN}) presents the Wightman function in the region
between the branes in the form of the image sum. In the spacetime region $%
\left( {{x}_{-}^{1}}\right) ^{2}+|\Delta \mathbf{x}|^{2}+\Delta z^{2}>\Delta
t^{2}$ one has the relation $u_{0,0}^{(-)}=\cosh \left( \sigma (x,x^{\prime
})/\alpha \right) $ with $\sigma (x,x^{\prime })$ being the geodesic
distance between the spacetime points $x$ and $x^{\prime }$. In the cases of
Dirichlet and Neumann boundary conditions, the Wightman function for the
geometry of a single brane at $x^{1}=a_{j}$ is obtained from (\ref{WDN}) in
the limit $(-1)^{j^{\prime }}a_{j^{\prime }}\rightarrow +\infty $. In the
series over $n$ the contribution of the term $n=1$, $j=j^{\prime }$ survives
only and we get the result obtained in \cite{Beze15}:%
\begin{equation}
W_{j}(x,x^{\prime })=W_{0}(x,x^{\prime })-\frac{\delta _{\mathrm{J}}\alpha
^{1-D}}{2^{\frac{D}{2}+\nu +1}\pi ^{\frac{D}{2}}}f_{\nu }\left( 1+\frac{%
\left\vert {{{x}_{+}^{1}}}-2a_{j}\right\vert ^{2}+|\Delta \mathbf{x}%
|^{2}+\Delta z^{2}-\Delta t^{2}}{2zz^{\prime }}\right) .  \label{WjDN}
\end{equation}

We can also consider the problem with Dirichlet boundary condition on the
brane $x^{1}=a_{1}$ and Neumann condition on the second brane. In this case $%
c_{j}(\lambda )=(-1)^{j}$ and the Wightman function is obtained in a way
similar to the cases of Dirichlet and Neumann conditions on both the branes.
It can be seen that the corresponding expression is obtained from (\ref{WDN}%
) by the replacements (the replacement of $\delta _{\mathrm{J}}$ should be
made after the summation sign $\sum_{j=1,2}$)%
\begin{equation}
\sum_{n=1}^{\infty }\rightarrow \sum_{n=1}^{\infty }(-1)^{n},\;\delta _{%
\mathrm{J}}\rightarrow (-1)^{j-1}.  \label{DNrepl}
\end{equation}%
In the regions $x^{1}<a_{1}$ and $x^{1}>a_{2}$ the Wightman functions for
the Dirichlet-Neumann combination of boundary conditions are given by (\ref%
{WjDN}) with $\delta _{\mathrm{J}}=1$ and $\delta _{\mathrm{J}}=-1$,
respectively.

\section{VEV of the field squared}

\label{sec:phi2}

In this section we investigate the VEV of the field squared $\langle
0|\varphi ^{2}|0\rangle \equiv \langle \varphi ^{2}\rangle $. It is obtained
taking the coincidence limit of the arguments in the Wightman function. Of
course, that limit is divergent and a renormalization procedure is required
to extract finite physical values. Here we are interested in the effects
induced by the branes. For points outside the branes the local geometry is
the same as that for AdS spacetime without branes. The divergences in the
coincidence limit are determined by the local geometrical characteristics
and we conclude that for $x^{1}\neq a_{j}$, $j=1,2$, they are the same as in
AdS spacetime. Having extracted the part in the Wightman function
corresponding to the latter geometry (the function $W_{0}(x,x^{\prime })$),
the renormalization is reduced to that for brane-free AdS spacetime. That
procedure for the VEVs of the field squared and of the energy-momentum
tensor is well investigated in the literature.

Taking the coincidence limit $x^{\prime }\rightarrow x$ in (\ref{Wf4}), the
VEV of the field squared is presented as%
\begin{eqnarray}
\langle \varphi ^{2}\rangle &=&\langle \varphi ^{2}\rangle _{0}+\frac{%
2^{2-D}\alpha ^{1-D}z^{D}}{\pi ^{\frac{D}{2}}\Gamma \left( \frac{D}{2}%
-1\right) }\int_{0}^{\infty }dk\,\,k^{D-3}\int_{0}^{\infty }d\gamma \,\gamma
J_{\nu }^{2}(\gamma z)  \notag \\
&&\times \int_{b}^{\infty }d\lambda \,\frac{2{+}\sum_{j=1,2}{e}^{2{|{{x}^{1}-%
{a_{j}}}|\lambda }}c_{j}(\lambda )}{\left[ c_{1}(\lambda )c_{2}(\lambda
)e^{2a\lambda }-1\right] \sqrt{\lambda ^{2}-b^{2}}},  \label{ph2}
\end{eqnarray}%
where $\langle \varphi ^{2}\rangle _{0}$ is the renormalized VEV in AdS
spacetime when the branes are absent. Because of the maximal symmetry of AdS
geometry the part $\langle \varphi ^{2}\rangle _{0}$ does not depend on the
spacetime point and it is well investigated in the literature. For further
transformation of the brane-induced contribution in (\ref{ph2}), instead of $%
\lambda $ we introduce a new integration variable $\chi =$ $\sqrt{\lambda
^{2}-b^{2}}$ and then pass to polar coordinates in the plane $(k,\chi )$.
After integrating over the angular part one finds%
\begin{equation}
\langle \varphi ^{2}\rangle =\langle \varphi ^{2}\rangle _{0}+\frac{\left( 2%
\sqrt{\pi }\alpha \right) ^{1-D}z^{D}}{\Gamma \left( \frac{D-1}{2}\right) }%
\int_{0}^{\infty }dr\,r^{D-2}\,\int_{0}^{\infty }d\gamma \,\frac{\gamma }{%
\lambda }\frac{2{+}\sum_{j=1,2}{e}^{2{|{{x}^{1}-{a_{j}}}|\lambda }%
}c_{j}(\lambda )}{c_{1}(\lambda )c_{2}(\lambda )e^{2a\lambda }-1}J_{\nu
}^{2}(\gamma z),  \label{ph22}
\end{equation}%
where $\lambda =\sqrt{\gamma ^{2}+r^{2}}$. Introducing polar coordinates in
the plane $(r,\gamma )$, for the angular integral we use the result \cite%
{Prud2}%
\begin{equation}
\int_{0}^{1}dxx(1-x^{2})^{\mu -3/2}J_{\nu }^{2}(ux)=\frac{\Gamma (\mu -1/2)}{%
2^{2\nu +1}}u^{2\nu }F_{\nu }^{\mu }(u),  \label{IntJ}
\end{equation}%
with the function%
\begin{equation}
F_{\nu }^{\mu }(u)=\frac{\,_{1}F_{2}(\nu +\frac{1}{2};\mu +\nu +\frac{1}{2}%
,1+2\nu ;-u^{2})}{\Gamma (\mu +\nu +\frac{1}{2})\Gamma (1+\nu )}.
\label{Fnu}
\end{equation}%
Here, $_{1}F_{2}(a;b,c;z)$ is the hypergeometric function. The final
expression reads%
\begin{equation}
\langle \varphi ^{2}\rangle =\langle \varphi ^{2}\rangle _{0}+\frac{\left(
\sqrt{\pi }\alpha \right) ^{1-D}}{2^{D+2\nu }}\int_{0}^{\infty
}dx\,x^{D+2\nu -1}F_{\nu }^{D/2}(x)\frac{2{+}\sum_{j=1,2}{e}^{2{|{{x}^{1}-{%
a_{j}}}|x/z}}c_{j}(x/z)}{c_{1}(x/z)c_{2}(x/z)e^{2ax/z}-1}.  \label{ph23}
\end{equation}

In a similar way, by making use of the formula (\ref{Wf3}), we can obtain
the representation%
\begin{equation}
\langle \varphi ^{2}\rangle =\langle \varphi ^{2}\rangle _{j}+\frac{\left(
\sqrt{\pi }\alpha \right) ^{1-D}}{2^{D+2\nu }}\int_{0}^{\infty
}dx\,x^{D+2\nu -1}F_{\nu }^{D/2}(x)\frac{2{+}\sum_{l=\pm 1}\left[ {e}^{2{|{{x%
}^{1}-{a_{j}}}|}x/z}c_{j}(x/z)\right] ^{l}}{c_{1}(x/z)c_{2}(x/z)e^{2ax/z}-1},
\label{ph24}
\end{equation}%
where the VEV in the geometry of a single brane at $x^{1}=a_{j}$ is
expressed as (see \cite{Beze15})%
\begin{equation}
\langle \varphi ^{2}\rangle _{j}=\langle \varphi ^{2}\rangle _{0}+\frac{%
\left( \sqrt{\pi }\alpha \right) ^{1-D}}{2^{D+2\nu }}\int_{0}^{\infty
}dx\,x^{D+2\nu -1}F_{\nu }^{D/2}(x)\frac{e^{-2{|{{x}^{1}-{a_{j}}}|x/z}}}{%
c_{j}({x/z})}.  \label{ph2j}
\end{equation}%
Note that the product $\alpha ^{D-1}\langle \varphi ^{2}\rangle $ depends on
the quantities having dimension of length ($x^{1}$, $a_{j}$, $\beta _{j}$)
and on the coordinate $z$ through the ratios $x^{1}/z$, $a_{j}/z$, $\beta
_{j}/z$. Those ratios are the proper values of the quantities measured by an
observer with fixed $z$ in units of the curvature radius $\alpha $. This
feature is a consequence of the AdS maximal symmetry.

For a conformally coupled massless field one has $\nu =1/2$ and%
\begin{equation}
F_{1/2}^{\mu }(u)=\frac{2}{\sqrt{\pi }u^{2}}\left[ \frac{1}{\Gamma \left(
\mu \right) }-\frac{J_{\mu -1}(2u)}{u^{\mu -1}}\right] .  \label{Fcc}
\end{equation}%
For the VEV of the field squared this gives%
\begin{equation}
\langle \varphi ^{2}\rangle =\langle \varphi ^{2}\rangle _{0}+\left(
z/\alpha \right) ^{D-1}\langle \varphi ^{2}\rangle _{\mathrm{(M)}},
\label{ph2cc}
\end{equation}%
where%
\begin{equation}
\langle \varphi ^{2}\rangle _{\mathrm{(M)}}=\frac{1}{2^{D}\pi ^{\frac{D}{2}}}%
\int_{0}^{\infty }d\lambda \,\lambda ^{D-2}\left[ \frac{1}{\Gamma (D/2)}-%
\frac{J_{D/2-1}(2z\lambda )}{(z\lambda )^{D/2-1}}\right] \frac{2{+}%
\sum_{j=1,2}{e}^{2{|{{x}^{1}-{a_{j}}}|\lambda }}c_{j}(\lambda )}{%
c_{1}(\lambda )c_{2}(\lambda )e^{2a\lambda }-1}.  \label{ph2Mcc}
\end{equation}%
The background geometry under consideration is conformally flat and the last
term in (\ref{ph2cc}) exhibits the standard conformal relation between the
boundary-induced parts of the VEVs in two conformally related problems (see,
for example, \cite{Birr82}). The geometry of two branes in AdS spacetime is
conformally connected to the problem in the Minkowski spacetime with the
line element%
\begin{equation}
ds_{\mathrm{M}}^{2}=dt^{2}-\left( dx^{1}\right) ^{2}-d\mathbf{x}^{2}-dz^{2},
\label{ds2M}
\end{equation}%
involving two parallel Robin plates at $x^{1}=a_{1}$ and $x^{1}=a_{2}$
intersected by the plate $z=0$ with Dirichlet boundary condition. The latter
plate is the conformal image of the AdS boundary. Note that the part in (\ref%
{ph2Mcc}) coming from the first term in the square brackets gives the mean
field squared in the region between two parallel plates in the Minkowski
spacetime (the boundary at $z=0$ is absent) and the part with the second
term is induced by the Dirichlet plate at $z=0$.

Note that the Dirichlet boundary condition at $z=0$ in the conformally
related problem on the Minkowski bulk is related to the condition we have
imposed for the scalar modes (\ref{mf}) on the AdS boundary. For the values
of the parameter $\nu $ in the range $0\leq \nu <1$ the general normalizable
solution of the field equation has the form (\ref{mf}) with the Bessel
function replaced by the linear combination $J_{\nu }(\gamma z)+b_{\sigma
}Y_{\nu }(\gamma z)$, where $Y_{\nu }(x)$ is the Neumann function. In this
case an additional boundary condition is required on the AdS boundary for
unique fixation of the set of modes. Our choice in (\ref{mf}) corresponds to
Dirichlet condition. In the literature the Neumann and more general Robin
boundary conditions have been considered as well (for recent discussions see
\cite{Ishi04}). In the conformally related problem on the Minkowski
spacetime, the boundary condition on the $z=0$ image is determined by the
respective condition on the AdS boundary. Note that the different boundary
conditions will correspond to different conformal field theories in the
context of the AdS/CFT correspondence.

Let us consider the Minkowskian limit of the problem at hand. It corresponds
to the limit $\alpha \rightarrow \infty $ for fixed value of the coordinate $%
y$ in (\ref{ds}). Introducing in (\ref{ph23}) a new integration variable $%
\lambda =x/z$ and by taking into account that in the limit under
consideration $z\approx \alpha $ and $\nu \approx m\alpha $, we see that
both the argument and the order $\nu $ of the function $F_{\nu
}^{D/2}(\lambda z)$ are large. The uniform asymptotic expansion is obtained
in \cite{Beze15} by using the corresponding expansion for the Bessel
function in (\ref{IntJ}). It has been shown that for large $\nu $ and $%
\lambda <m$ the function $F_{\nu }^{D/2}(\nu \lambda /m)$ is exponentially
small. The VEV of the field squared is dominated by the contribution of the
integral coming from the region $\lambda >m$. In that region the leading
term in the expansion over $1/\nu $ is given by \cite{Beze15}%
\begin{equation}
F_{\nu }^{\mu }\left( \frac{\nu }{m}\lambda \right) \approx \frac{\left(
\lambda ^{2}-m^{2}\right) ^{\mu -1}(2m/\nu )^{2\nu +1}}{2\sqrt{\pi }\Gamma
(D/2)\lambda ^{2\mu +2\nu -1}}.  \label{FnuLarge2}
\end{equation}%
With this estimate we get $\lim_{\alpha \rightarrow \infty }\langle \varphi
^{2}\rangle =\langle \varphi ^{2}\rangle _{\mathrm{(M)}}^{(0)}$, where%
\begin{equation}
\langle \varphi ^{2}\rangle _{\mathrm{(M)}}^{(0)}=\frac{\left( 4\pi \right)
^{-\frac{D}{2}}}{\Gamma (D/2)}\int_{m}^{\infty }d\lambda \,\left( \lambda
^{2}-m^{2}\right) ^{D/2-1}\frac{2{+}\sum_{j=1,2}{e}^{2{|{{x}^{1}-{a_{j}}}%
|\lambda }}c_{j}(\lambda )}{c_{1}(\lambda )c_{2}(\lambda )e^{2a\lambda }-1},
\label{ph2M0m}
\end{equation}%
is the mean field squared in the region between two Robin plates in
background of Minkowski spacetime with the line element (\ref{ds2M}). This
result for a massive field was presented in \cite{SahaRev}. For a massless
field it is reduced to the result derived in \cite{Rome02}.

In order to find the mean field squared on AdS bulk in the special cases of
Dirichlet and Neumann boundary conditions we can use the representation (\ref%
{WDN}) for the Wightman function. The corresponding expression reads%
\begin{equation}
\langle \varphi ^{2}\rangle =\langle \varphi ^{2}\rangle _{0}+\frac{\alpha
^{1-D}}{2^{\frac{D}{2}+\nu +1}\pi ^{\frac{D}{2}}}\sum_{n=1}^{\infty }\left[
2f_{\nu }(u_{n}){-}\delta _{\mathrm{J}}\sum_{j=1,2}f_{\nu }(u_{j,n})\right] ,
\label{ph2DN}
\end{equation}%
with the notations%
\begin{eqnarray}
u_{n} &=&1+2\left( na/z\right) ^{2},  \notag \\
u_{j,n} &=&1+\frac{2}{z^{2}}\left( na-{|{{x}^{1}-{a_{j}}}|}\right) ^{2}.
\label{ujn}
\end{eqnarray}%
An alternative representation is obtained from (\ref{ph23}) expanding the
function $1/(e^{2ax/z}-1)$. The integral is evaluated by using the formula
from \cite{Prud3}:%
\begin{equation}
\int_{0}^{\infty }dx\,x^{2\mu +2\nu -1}e^{-2cx}F_{\nu }^{\mu }(x)=\frac{%
h_{\nu }^{\mu }(c)}{2\sqrt{\pi }},  \label{Int2}
\end{equation}%
where the function in the right-hand side is defined as%
\begin{equation}
h_{\nu }^{\mu }(u)=\frac{\Gamma \left( \mu +\nu \right) }{\Gamma (\nu +1)}%
\frac{1}{u^{2\left( \mu +\nu \right) }}F\left( \nu +\frac{1}{2},\mu +\nu
;1+2\nu ;-\frac{1}{u^{2}}\right) .  \label{hnu}
\end{equation}%
The VEV is presented as%
\begin{equation}
\langle \varphi ^{2}\rangle =\langle \varphi ^{2}\rangle _{0}+\frac{\alpha
^{1-D}}{2^{D+2\nu +1}\pi ^{\frac{D}{2}}}\sum_{n=1}^{\infty }\left[ 2h_{\nu
}^{\frac{D}{2}}\left( \frac{na}{z}\right) {-}\delta _{\mathrm{J}%
}\sum_{j=1,2}h_{\nu }^{\frac{D}{2}}\left( \frac{na-{|{{x}^{1}-{a_{j}}}|}}{z}%
\right) \right] .  \label{ph2DN2}
\end{equation}%
By employing the linear and quadratic transformation formulas for the
hypergeometric function (see, for example, \cite{Abra}) we can see that%
\begin{equation}
h_{\nu }^{\frac{D}{2}}(x)=2^{\frac{D}{2}+\nu }f_{\nu }(1+2x^{2}).
\label{hfrel}
\end{equation}%
This relation shows the equivalence of the representations (\ref{ph2DN}) and
(\ref{ph2DN2}). Note that for a conformally coupled massless field%
\begin{equation}
h_{1/2}^{\mu }(x)=\frac{4\Gamma \left( \mu +1/2\right) }{\sqrt{\pi }\left(
2\mu -1\right) }\left[ x^{1-2\mu }-\left( x^{2}+1\right) ^{\frac{1}{2}-\mu }%
\right] .  \label{hcc}
\end{equation}

For Dirichlet or Neumann boundary conditions, the VEV in the problem with a
single brane is obtained from (\ref{ph2DN2}) taking the limit $%
a_{1}\rightarrow -\infty $ or $a_{2}\rightarrow +\infty $:%
\begin{equation}
\langle \varphi ^{2}\rangle _{j}=\langle \varphi ^{2}\rangle _{0}-\frac{%
\delta _{\mathrm{J}}\alpha ^{1-D}}{2^{D+2\nu +1}\pi ^{\frac{D}{2}}}h_{\nu }^{%
\frac{D}{2}}\left( \frac{\left\vert {{{x}^{1}}}-a_{j}\right\vert }{z}\right)
.  \label{ph2jDN}
\end{equation}%
In problems with two scalar fields with Dirichlet and Neumann conditions on
a single brane, the brane-induced mean field squared vanishes as a result of
cancelations of contributions from Dirichlet and Neumann scalars. In
particular, for $D=3$, the electromagnetic field with perfectly conducting
boundary condition on the brane is reduced to two scalar modes with
Dirichlet and Neumann conditions and their contributions in the vacuum
energy density cancel each other. An equivalent representation for the
single brane mean field squared $\langle \varphi ^{2}\rangle _{j}$, given in
\cite{Beze15}, is derived from (\ref{WjDN}) in the coincidence limit. In (%
\ref{ph2DN}) and (\ref{ph2DN2}), the parts corresponding to the contribution
of the brane at $x^{1}=a_{j^{\prime }\text{,}}$ when the second brane is
absent, are presented by the term $n=1$, $j=2$ for $j^{\prime }=1$ and by
the term $n=1$, $j=1$ for $j^{\prime }=2.$

In the case of Dirichlet boundary condition on the brane $x^{1}=a_{1}$ and
Neumann boundary condition on $x^{1}=a_{2}$ the expression for the mean
field squared in the region between the branes is obtained from (\ref{ph2DN2}%
) making the replacements (\ref{DNrepl}).

Now let us consider the behavior of the VEV $\langle \varphi ^{2}\rangle $
in asymptotic regions of the parameters. The VEV diverges on the branes. The
divergences come from the single brane contributions: in the representation (%
\ref{ph24}) the divergence at $x^{1}=a_{j}$ is contained in the part $%
\langle \varphi ^{2}\rangle _{j}$ (in the last term of (\ref{ph2j})). Near
the brane, for $\left\vert {{{x}^{1}}}-a_{j}\right\vert \ll z$, the total
VEV $\langle \varphi ^{2}\rangle $ is dominated by the last term in (\ref%
{ph2j}). Assuming additionally $\left\vert {{{x}^{1}}}-a_{j}\right\vert \ll
|\beta _{j}|$ (non-Dirichlet boundary conditions), the leading term in the
expansion over the distance from the brane reads \cite{Beze15}%
\begin{equation}
\langle \varphi ^{2}\rangle \approx \frac{\Gamma \left( \frac{D-1}{2}\right)
}{\left( 4\pi \right) ^{\frac{D+1}{2}}}\left( \frac{z}{\alpha \left\vert {{{x%
}^{1}}}-a_{j}\right\vert }\right) ^{D-1}.  \label{ph2near}
\end{equation}%
For the Dirichlet boundary condition the corresponding asymptotic differs
from (\ref{ph2near}) by the sign of the right-hand side. The last term in
the representation (\ref{ph24}) is finite on the brane ${{{x}^{1}}}=a_{j}$.

For points near the AdS boundary, $z\ll |x^{1}-a_{j}|$, $j=1,2$, the main
contribution to the integral in (\ref{ph23}) comes from the region with
small values of $x$. By using the asymptotic expression $F_{\nu }^{\mu
}(x)\approx F_{\nu }^{\mu }(0)\left( 1+\mathcal{O}(x^{2})\right) $ with%
\begin{equation}
F_{\nu }^{\mu }(0)=\frac{1}{\Gamma \left( \nu +1\right) \Gamma \left( \mu
+\nu +\frac{1}{2}\right) },  \label{Fmusmall}
\end{equation}%
in the leading order, for the brane-induced contribution we get%
\begin{equation}
\langle \varphi ^{2}\rangle \approx \langle \varphi ^{2}\rangle _{0}+\frac{%
F_{\nu }^{\frac{D}{2}}(0)z^{D+2\nu }}{2^{D+2\nu }\left( \sqrt{\pi }\alpha
\right) ^{D-1}}\int_{0}^{\infty }d\lambda \,\lambda ^{D+2\nu -1}\frac{2{+}%
\sum_{j=1,2}{e}^{2{|{{x}^{1}-{a_{j}}}|\lambda }}c_{j}({\lambda })}{c_{1}({%
\lambda })c_{2}({\lambda })e^{2a{\lambda }}-1}.  \label{ph2nearAdSb}
\end{equation}%
Hence, for points near the AdS boundary and not too close to the branes, the
brane-induced part in the mean field squared tends to zero like $z^{D+2\nu }$%
. For points near the horizon, $z\gg a$, the integral in (\ref{ph23}) is
dominated by the contribution coming from the region with large values of $x$%
. For those $x$ one has \cite{Beze15}
\begin{equation}
F_{\nu }^{\mu }(x)\approx \frac{2^{2\nu }}{\sqrt{\pi }\Gamma \left( \mu
\right) x^{2\nu +1}},\;x\gg 1,  \label{Fmularge}
\end{equation}%
and the VEV of the field squared is approximated by
\begin{equation}
\langle \varphi ^{2}\rangle \approx \langle \varphi ^{2}\rangle _{0}+\left(
z/\alpha \right) ^{D-1}\langle \varphi ^{2}\rangle _{\mathrm{(M)}%
}^{(0)}|_{m=0},  \label{ph2nearH}
\end{equation}%
where $\langle \varphi ^{2}\rangle _{\mathrm{(M)}}^{(0)}|_{m=0}$ (see (\ref%
{ph2M0m})) is the corresponding VEV for a massless scalar field between two
parallel plates in the Minkowski bulk with separation $a$ \cite{Rome02}.
Note that the latter is obtained from (\ref{ph2Mcc}) in the limit $%
z\rightarrow \infty $. As seen, near the horizon the effects of the
curvature on the brane-induced VEV are weak. Note that for a given $a$ and
large $z$ the proper separation between the branes is much smaller than the
curvature radius, $a_{p}=\alpha a/z\ll \alpha $, and the main contribution
to the brane-induced VEV comes from the vacuum fluctuations with the
wavelengths much smaller than the curvature radius. The influence of the
gravitational field on those fluctuations is weak.

\begin{figure}[tbph]
\begin{center}
\begin{tabular}{cc}
\epsfig{figure=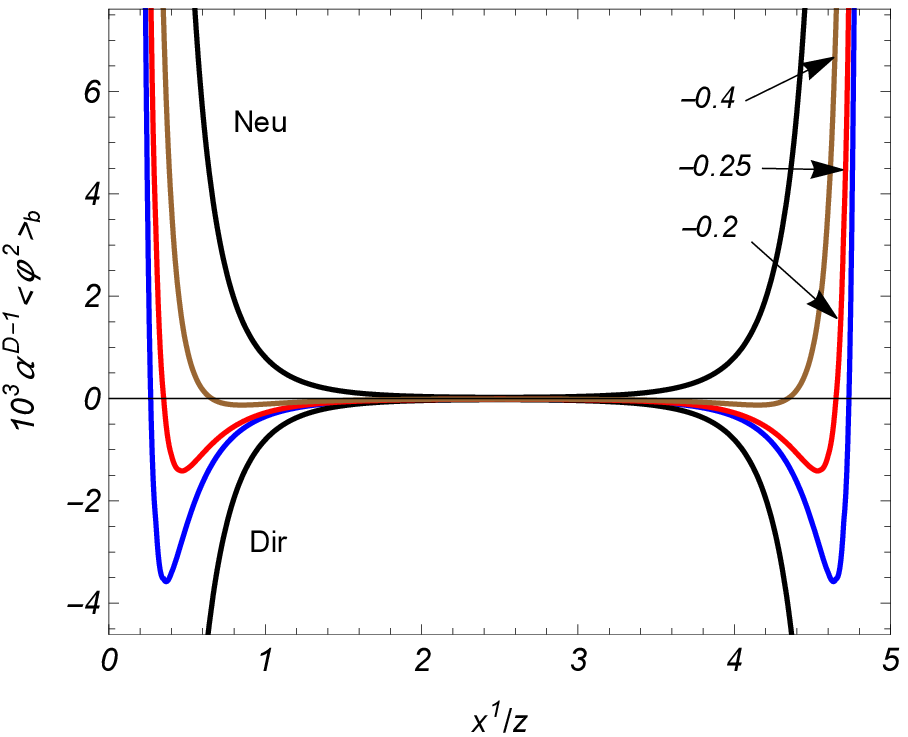,width=7.5cm,height=6cm} & \quad %
\epsfig{figure=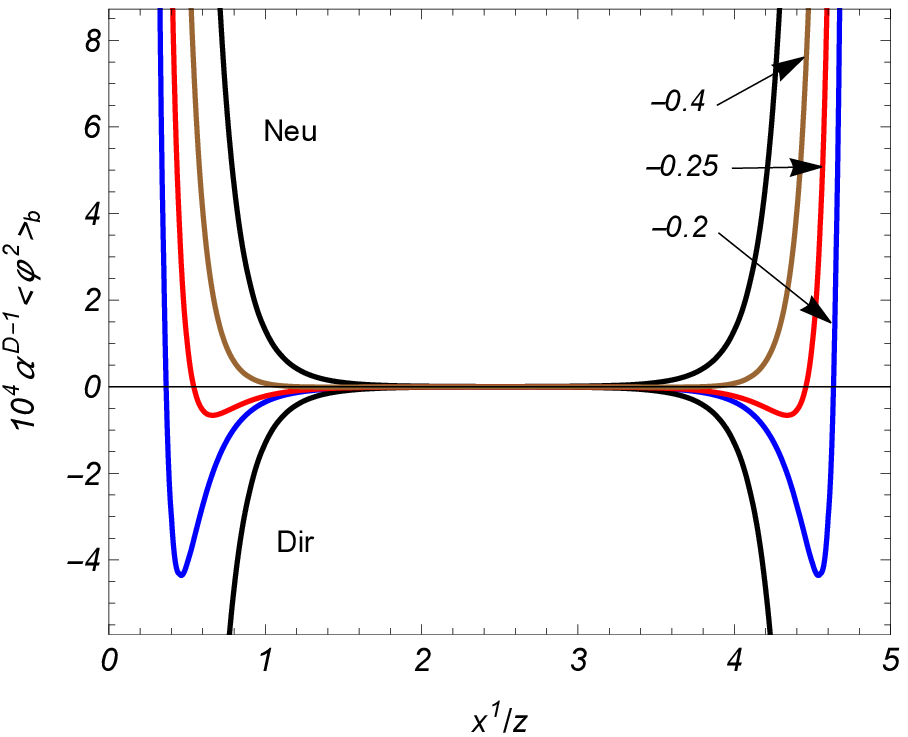,width=7.5cm,height=6cm}%
\end{tabular}%
\end{center}
\caption{The brane-induced mean field squared in the region between the
branes as a function of the ratio $x^{1}/z$ for $D=4$ fields with conformal
and minimal couplings (left and right panels, respectively). The graphs are
plotted for $m\protect\alpha =0.5$ and the numbers near the curves
correspond to the values of the ratio $\protect\beta _{1}/z=\protect\beta %
_{2}/z$. The graphs for Robin boundary conditions are located in the region
between the curves corresponding to Dirichlet and Neumann conditions.}
\label{fig2}
\end{figure}

In Figure \ref{fig2} the brane-induced VEV of the field squared, $\langle
\varphi ^{2}\rangle _{\mathrm{b}}=\langle \varphi ^{2}\rangle -\langle
\varphi ^{2}\rangle _{0}$, is plotted in the region between the branes as a
function of the proper distance from the brane at $x^{1}=0$ (in units of the
AdS curvature radius $\alpha $). For the location of the second brane we
have taken $a_{2}/z=5$. The graphs are plotted for $D=4$ conformally (left
panel) and minimally (right panel) coupled massive scalar fields with $%
m\alpha =0.5$. The same boundary conditions are imposed on the branes ($%
\beta _{1}=\beta _{2}$) and the numbers near the curves indicate the
respective values of the ratio $\beta _{1}/z$. The graphs for Dirichlet and
Neumann boundary conditions are presented as well (Dir and Neu,
respectively). The brane-induced mean field squared is negative for the
Dirichlet case and positive for the Neumann one. For Robin conditions with
sufficiently small values of $|\beta _{j}|/z$, the VEV is positive near the
branes and negative in the region near the center with respect to the
branes. With increasing $|\beta _{j}|/z$, started from some critical value,
the VEV $\langle \varphi ^{2}\rangle _{\mathrm{b}}$ becomes positive
everywhere in the region between the branes. For the example presented in
Figure \ref{fig2}, for the critical values one has $\beta _{j}/z\approx
-1.08 $ and $\beta _{j}/z\approx -0.70$ for conformally and minimally
coupled scalars, respectively. The critical value for $|\beta _{j}|/z$
decreases with decreasing $a/z$.

\section{Energy-momentum tensor}

\label{sec:EMT}

Another important characteristic of the vacuum state is the VEV\ of the
energy-momentum tensor. With the known Wightman function and the VEV of the
field squared, it is evaluated by using the formula
\begin{equation}
\langle T_{ik}\rangle =\frac{1}{2}\lim_{x^{\prime }\rightarrow x}\left( {%
\partial _{i}\partial _{k}^{\prime }+\partial _{k}\partial _{i}^{\prime }}%
\right) W(x,x^{\prime })+\hat{B}_{ik}\langle \varphi ^{2}\rangle ,
\label{Tik}
\end{equation}%
where the operator acting on the VEV of the field squared is defined by
\begin{equation}
\hat{B}_{ik}=\left( \xi -\frac{1}{4}\right) g_{ik}g^{lm}\nabla _{l}\nabla
_{m}-\xi \left( \nabla _{i}\nabla _{k}+R_{ik}\right) ,  \label{Bik}
\end{equation}%
with $R_{ik}=-Dg_{ik}/\alpha ^{2}$ being the Ricci tensor for AdS spacetime.
In the geometry at hand one gets%
\begin{equation}
\hat{B}_{00}=\left( \frac{1}{4}-\xi \right) \left( \partial
_{1}^{2}+\partial _{z}^{2}-\frac{D-1}{z}\partial _{z}\right) -\frac{\xi }{z}%
\partial _{z}+\frac{D}{z^{2}}\xi ,  \label{B00}
\end{equation}%
and the spatial diagonal components are expressed as $\hat{B}_{ll}=-\hat{B}%
_{00}-\hat{C}_{ll}$, $l=1,2,\ldots ,D$, where%
\begin{eqnarray}
\hat{C}_{11} &=&\xi \partial _{1}^{2},\;\hat{C}_{ll}=0,\;l=2,\ldots ,D-1,
\notag \\
\hat{C}_{DD} &=&\xi \left( \partial _{z}^{2}+\frac{2}{z}\partial _{z}\right)
.  \label{CDD}
\end{eqnarray}%
In addition to the diagonal components, the action of the operator (\ref{Bik}%
) on $\langle \varphi ^{2}\rangle $ gives a nonzero off-diagonal component $%
\hat{B}_{1D}\langle \varphi ^{2}\rangle $ with the operator%
\begin{equation}
\hat{B}_{1D}=-\xi \left( \partial _{z}+\frac{1}{z}\right) \partial _{1}.
\label{B1D}
\end{equation}

By using the representation (\ref{Wf4}), the coincidence limit of the
bitensor ${\partial _{i}\partial _{k}^{\prime }}W(x,x^{\prime })$ is
evaluated in a way we have described above for the mean field squared. For
the diagonal components $\langle T_{ll}\rangle $ with $l\neq D$ the angular
integrals at the last step are expressed in terms of the functions $F_{\nu
}^{D/2}(u)$ and $F_{\nu }^{D/2+1}(u)$. For the component $\langle
T_{DD}\rangle $ the integral is reduced to
\begin{equation}
\int_{0}^{1}dx\,x\left( 1-x^{2}\right) ^{\frac{D-3}{2}}\left[ \partial
_{u}\left( u^{\frac{D}{2}}J_{\nu }(ux)\right) \right] ^{2}.  \label{Int}
\end{equation}%
By making use of the equation for the Bessel function this integral is
expressed in terms of the functions $F_{\nu }^{D/2}(u)$, $F_{\nu
}^{D/2+1}(u) $ and of the first and second derivatives of $F_{\nu }^{D/2}(u)$%
.

After long but straightforward calculations, the VEVs of the diagonal
components of the energy-momentum tensor are written in the form (no
summation over $i$)%
\begin{eqnarray}
\langle T_{i}^{i}\rangle &=&\langle T_{i}^{i}\rangle _{0}-\frac{\alpha
^{-1-D}}{2^{D+2\nu }\pi ^{\frac{D-1}{2}}}\mathbf{\,}\int_{0}^{\infty
}dx\,x\left\{ \frac{E_{i}x^{D+2\nu }F_{\nu }^{\frac{D}{2}}(x)}{%
c_{1}(x/z)c_{2}(x/z)e^{2ax/z}-1}\right.  \notag \\
&&\left. +\frac{2{+}\sum_{j=1,2}{e}^{2{|{{x}^{1}-{a_{j}}}|x/z}}c_{j}(x/z)}{%
c_{1}(x/z)c_{2}(x/z)e^{2ax/z}-1}\left[ A_{i}x^{D+2\nu }F_{\nu }^{\frac{D}{2}%
+1}(x)+\hat{B}_{i}x^{D+2\nu }F_{\nu }^{\frac{D}{2}}(x)\right] \right\} .
\label{Tii}
\end{eqnarray}%
Here we have defined the operators%
\begin{eqnarray}
\hat{B}_{1} &=&\left( \xi -\frac{1}{4}\right) \partial _{x}^{2}+\left[ \frac{%
D-1}{4}-\left( D-2\right) \xi \right] \frac{\partial _{x}}{x}-\frac{D\xi }{%
x^{2}},  \notag \\
\hat{B}_{i} &=&\hat{B}_{1}+4\xi -1,\;i=0,2,\ldots ,D-1,  \notag \\
\hat{B}_{D} &=&\frac{1}{4}\partial _{x}^{2}-D\left( \xi +\xi _{D}\right)
\frac{\partial _{x}}{x}+\frac{D^{2}\xi -m^{2}\alpha ^{2}}{x^{2}}+4\xi ,
\label{BD}
\end{eqnarray}%
and the coefficients%
\begin{eqnarray}
E_{i} &=&2\left( 1-4\xi \right) ,\;i=0,2,\ldots ,D,\;E_{1}=-2,  \notag \\
A_{i} &=&\frac{1}{2},\;i=0,2,\ldots ,D-1,\;A_{D}=\frac{1-D}{2},  \label{Ai}
\end{eqnarray}%
and $A_{1}=0$. The nonzero off-diagonal component is expressed as%
\begin{equation}
\langle T_{D}^{1}\rangle =-\frac{2\alpha ^{-1-D}}{2^{D+2\nu }\pi ^{\frac{D-1%
}{2}}}\mathbf{\,}\int_{0}^{\infty }dx\frac{\sum_{j=1,2}\left( -1\right) ^{j}{%
e}^{2{|{{x}^{1}-{a_{j}}}|}x/z}c_{j}(x/z)}{c_{1}(x/z)c_{2}(x/z)e^{2ax/z}-1}%
\left[ \left( \xi -\frac{1}{4}\right) x\partial _{x}+\xi \right] x^{D+2\nu
}F_{\nu }^{\frac{D}{2}}(x).  \label{T1D}
\end{equation}%
In (\ref{Tii}), the part $\langle T_{i}^{k}\rangle _{0}$ corresponds to the
vacuum energy-momentum tensor in the brane-free AdS spacetime. Similar to
the case of the VEV $\langle \varphi ^{2}\rangle _{0}$, that part is
well-known from the literature. From the maximal symmetry of the AdS
geometry one has $\langle T_{i}^{k}\rangle _{0}=\mathrm{const}\cdot \delta
_{i}^{k}$. The components $\langle T_{0}^{0}\rangle $ and $\langle
T_{i}^{i}\rangle $, $i=2,\ldots ,D-1$, determining the energy density and
stresses along the directions parallel to the branes (except the component $%
i=D$), are equal. Of course, that is a consequence of the problem symmetry.
As another consequence of the symmetry, the VEV of the energy-momentum
tensor depends on the variables $x^{1}$, $a_{j}$, $\beta _{j}$, $z$ in terms
of the combinations $x^{1}/z$, $a_{j}/z$, $\beta _{j}/z$. The first and
second derivatives of the product $x^{D+2\nu }F_{\nu }^{D/2}(x)$, appearing
in (\ref{Tii}) and (\ref{T1D}), are expressed in terms of the functions $%
F_{\nu }^{D/2}(x)$, $F_{\nu }^{D/2-1}(x)$, and $F_{\nu }^{D/2-2}(x)$. The
corresponding relations can be found in \cite{Beze15}.

Let us denote by $\langle T_{i}^{k}\rangle _{\mathrm{b}}=\langle
T_{i}^{k}\rangle -\langle T_{i}^{k}\rangle _{0}$ the brane-induced
contribution to the vacuum energy-momentum tensor. We can check the
following relation for the corresponding trace:
\begin{equation}
\langle T_{i}^{i}\rangle _{\mathrm{b}}=D\left( \xi -\xi _{D}\right) \nabla
_{l}\nabla ^{l}\langle \varphi ^{2}\rangle _{\mathrm{b}}+m^{2}\langle
\varphi ^{2}\rangle _{\mathrm{b}},  \label{Trace}
\end{equation}%
where the brane-induced part in the VEV of the field squared is given by the
last term in (\ref{ph23}). The trace is zero for a conformally coupled
massless field. Another relation expected from general arguments is the
covariant conservation equation $\nabla _{k}\langle T_{i}^{k}\rangle _{%
\mathrm{b}}=0$. The latter is a necessary condition for $\langle
T_{i}^{k}\rangle _{\mathrm{b}}$ to be a source in the Einstein field
equations. From the equations with $i=1$ and $i=D$ the following two
relations are obtained between the separate components (see also \cite%
{Beze15} for the corresponding relations in the geometry of a single brane):%
\begin{eqnarray}
\partial _{1}\langle T_{1}^{1}\rangle _{\mathrm{b}} &=&-z^{D+1}\partial
_{z}\left( \frac{\langle T_{1}^{D}\rangle _{\mathrm{b}}}{z^{D+1}}\right) ,
\notag \\
\partial _{1}\langle T_{D}^{1}\rangle _{\mathrm{b}} &=&-z^{D}\partial
_{z}\left( \frac{\langle T_{D}^{D}\rangle _{\mathrm{b}}}{z^{D}}\right) -%
\frac{1}{z}\sum_{k=0}^{D-1}\langle T_{k}^{k}\rangle _{\mathrm{b}}.
\label{Eqcons}
\end{eqnarray}%
The first equation shows that the dependence of the normal stress on the
coordinate $x^{1}$ is related to the presence of the nonzero off-diagonal
component.

The VEV of the energy-momentum tensor in the geometry of a single brane at $%
x^{1}=a_{j}$ is obtained from (\ref{Tii}) and (\ref{T1D}) in the limit $%
(-1)^{j^{\prime }}a_{j^{\prime }}\rightarrow \infty $ with $j^{\prime }\neq
j $. For the diagonal components this gives (no summation over $i$)%
\begin{equation}
\langle T_{i}^{i}\rangle _{j}=\langle T_{i}^{i}\rangle _{0}-\frac{\alpha
^{-1-D}}{2^{D+2\nu }\pi ^{\frac{D-1}{2}}}\mathbf{\,}\int_{0}^{\infty }dx\,x%
\frac{{e}^{-2{|{{x}^{1}-{a_{j}}}|x/z}}}{c_{j}(x/z)}\left[ A_{i}x^{D+2\nu
}F_{\nu }^{\frac{D}{2}+1}(x)+\hat{B}_{i}x^{D+2\nu }F_{\nu }^{\frac{D}{2}}(x)%
\right] .  \label{Tij}
\end{equation}%
The corresponding expression for the off-diagonal component reads%
\begin{equation}
\langle T_{D}^{1}\rangle _{j}=\frac{2\left( -1\right) ^{j}\alpha ^{-1-D}}{%
2^{D+2\nu }\pi ^{\frac{D-1}{2}}}\mathbf{\,}\int_{0}^{\infty }dx\frac{{e}^{-2{%
|{{x}^{1}-{a_{j}}}|}x/z}}{c_{j}(x/z)}\left[ \left( \xi -\frac{1}{4}\right)
x\partial _{x}+\xi \right] x^{D+2\nu }F_{\nu }^{\frac{D}{2}}(x).
\label{T1Dj}
\end{equation}%
The formulae (\ref{Tij}) and (\ref{T1Dj}) were obtained in \cite{Beze15}
from the Wightman function (\ref{Wj}) by using (\ref{Tik}). Note that (\ref%
{T1Dj}) presents the VEV in the region $x^{1}>a_{1}$ for $j=1$ and in the
region $x^{1}<a_{2}$ for $j=2$. Making the replacement
\begin{equation}
\left( -1\right) ^{j}\rightarrow \mathrm{sgn}(a_{j}-x^{1}),  \label{replT1D}
\end{equation}%
in (\ref{T1Dj}) we obtain the expression for a single brane at $x^{1}=a_{j}$
that is valid for both regions $x^{1}<a_{j}$ and $x^{1}>a_{j}$.

Extracting the single brane contributions from the VEVs we can obtain the
following equivalent representations for the components of the vacuum
energy-momentum tensor (no summation over $i$):%
\begin{eqnarray}
\langle T_{i}^{i}\rangle &=&\langle T_{i}^{i}\rangle _{j}-\frac{\alpha
^{-1-D}}{2^{D+2\nu }\pi ^{\frac{D-1}{2}}}\mathbf{\,}\int_{0}^{\infty
}dx\,x\left\{ \frac{E_{i}x^{D+2\nu }F_{\nu }^{\frac{D}{2}}(x)}{%
c_{1}(x/z)c_{2}(x/z)e^{2ax/z}-1}\right.  \notag \\
&&\left. +\frac{2{+}\sum_{l=\pm 1}\left[ {e}^{2{|{{x}^{1}-{a_{j}}}|x/z}%
}c_{j}(x/z)\right] ^{l}}{c_{1}(x/z)c_{2}(x/z)e^{2ax/z}-1}\left[
A_{i}x^{D+2\nu }F_{\nu }^{\frac{D}{2}+1}(x)+\hat{B}_{i}x^{D+2\nu }F_{\nu }^{%
\frac{D}{2}}(x)\right] \right\} ,  \label{Tii2}
\end{eqnarray}%
and%
\begin{eqnarray}
\langle T_{D}^{1}\rangle &=&\langle T_{D}^{1}\rangle _{j}-\frac{\left(
-1\right) ^{j}\alpha ^{-1-D}}{2^{D+2\nu -1}\pi ^{\frac{D-1}{2}}}\mathbf{\,}%
\int_{0}^{\infty }dx\frac{\sum_{l=\pm 1}l\left[ {e}^{2{|{{x}^{1}-{a_{j}}}|}%
x/z}c_{j}(x/z)\right] ^{l}}{c_{1}(x/z)c_{2}(x/z)e^{2ax/z}-1}  \notag \\
&&\times \left[ \left( \xi -\frac{1}{4}\right) x\partial _{x}+\xi \right]
x^{D+2\nu }F_{\nu }^{\frac{D}{2}}(x).  \label{T1D2}
\end{eqnarray}%
The last terms in these representations are the contributions induced by the
brane at $x^{1}=a_{j^{\prime }}$ when we add it to the problem with a single
brane at $x^{1}=a_{j}$. Those terms are finite on the brane $x^{1}=a_{j}$
and the divergences on that brane come from the single brane contribution $%
\langle T_{i}^{k}\rangle _{j}$. For points near the brane the total VEV is
dominated by the single brane contribution. Under the conditions $\left\vert
{{{x}^{1}}}-a_{j}\right\vert \ll z,|\beta _{j}|$, the corresponding leading
terms in the expansion over the distance from the brane are given in \cite%
{Beze15}:%
\begin{equation}
\langle T_{0}^{0}\rangle _{\mathrm{b}}\approx \frac{\left( 1-D\right)
\langle T_{1}^{1}\rangle _{\mathrm{b}}}{\left( |x^{1}-a_{j}|/z\right) ^{2}}%
\approx \frac{z\langle T_{D}^{1}\rangle _{\mathrm{b}}}{x^{1}-a_{j}}\approx
\frac{2D\left( \xi _{D}-\xi \right) \Gamma \left( \frac{D+1}{2}\right) }{\pi
^{\frac{D+1}{2}}\left( 2\alpha |x^{1}-a_{j}|/z\right) ^{D+1}}.  \label{Tnear}
\end{equation}%
For Dirichlet boundary condition, $\beta _{j}=0$, the leading terms are
given by the same expressions with opposite signs. As seen, the divergence
on the branes is weaker for the normal stress and off-diagonal component.
For conformal coupling the leading terms vanish and the next terms in the
expansion should be kept.

In the case of a conformally coupled massless field, by using the expression
(\ref{Fcc}) for the function $F_{\nu }^{\mu }(x)$, the vacuum
energy-momentum tensor is presented in the form%
\begin{equation}
\langle T_{k}^{i}\rangle =\langle T_{k}^{i}\rangle _{0}+\left( z/\alpha
\right) ^{D+1}\langle T_{k}^{i}\rangle _{\mathrm{(M)}},  \label{Ticc}
\end{equation}%
where $\langle T_{i}^{k}\rangle _{\mathrm{(M)}}$ is the corresponding VEV in
the region $a_{1}<x^{1}<a_{2}$, $z>0$ for the geometry of plates at $%
x^{1}=a_{1},a_{2}$ and $z=0$ in the Minkowski spacetime with the line
element (\ref{ds2M}). On the plates $x^{1}=a_{1},a_{2}$ the field obeys
Robin boundary condition (\ref{bc2}) and on the plate $z=0$ the Dirichlet
boundary condition is imposed. For the diagonal components the Minkowskian
VEV is given by (no summation over $i$)
\begin{eqnarray}
\langle T_{i}^{i}\rangle _{\mathrm{M}} &=&\langle T_{i}^{i}\rangle _{\mathrm{%
M}}^{(0)}+\frac{\pi ^{-\frac{D}{2}}}{2^{D+1}D}\mathbf{\,}\int_{0}^{\infty
}d\lambda \,\frac{\lambda ^{D}}{c_{1}(\lambda )c_{2}(\lambda )e^{2a\lambda
}-1}\Bigg\{a^{(i)}g_{\frac{D}{2}-1}(\lambda z)  \notag \\
&&+\left[ 2{+}\sum_{j=1,2}{e}^{2{|{{x}^{1}-{a_{j}}}|\lambda }}c_{j}(\lambda )%
\right] \left[ b^{(i)}g_{\frac{D}{2}-1}(\lambda z)+c^{(i)}g_{\frac{D}{2}%
}(\lambda z)\right] \Bigg\},  \label{TiiM}
\end{eqnarray}%
with the coefficients%
\begin{eqnarray}
(a^{(0)},a^{(1)},a^{(D)}) &=&(4,-4D,4),  \notag \\
(b^{(0)},b^{(1)},b^{(D)}) &=&(0,2,-2),  \notag \\
(c^{(0)},c^{(1)},c^{(D)}) &=&(1,1-D,0).  \label{ci}
\end{eqnarray}%
In (\ref{TiiM}) we have introduced the function
\begin{equation}
g_{\mu }(x)=x^{-\mu }J_{\mu }(2x),  \label{gmu}
\end{equation}%
and (no summation over $i$)
\begin{equation}
\langle T_{i}^{k}\rangle _{\mathrm{M}}^{(0)}=-\delta _{i}^{k}\frac{%
(-D)^{\delta _{i}^{1}}(4\pi )^{-\frac{D}{2}}}{\Gamma \left( \frac{D}{2}%
+1\right) }\int_{0}^{\infty }d\lambda \,\frac{\lambda ^{D}}{c_{1}(\lambda
)c_{2}(\lambda )e^{2a\lambda }-1}  \label{TiiM0}
\end{equation}%
is the corresponding VEV in the problem where the plate $z=0$ is absent.
Hence, the last term in (\ref{TiiM}) is induced by the Dirichlet plate $z=0$
added to the geometry of two parallel plates. For the off-diagonal component
in the Minkowski bulk we obtain%
\begin{equation}
\langle T_{D}^{1}\rangle _{\mathrm{(M)}}=\frac{2z^{1-\frac{D}{2}}}{2^{D+2\nu
}\pi ^{\frac{D}{2}}D}\mathbf{\,}\int_{0}^{\infty }d\lambda \,\lambda ^{\frac{%
D}{2}+1}\frac{\sum_{j=1,2}\left( -1\right) ^{j}{e}^{2{|{{x}^{1}-{a_{j}}}|}%
\lambda }c_{j}(\lambda )}{c_{1}(\lambda )c_{2}(\lambda )e^{2a\lambda }-1}J_{%
\frac{D}{2}}(2\lambda z).  \label{T1DM}
\end{equation}

For the VEV of the energy-momentum tensor, the consideration of the
Minkowskian limit, corresponding to large values of the curvature radius $%
\alpha $, is similar to that for the mean field squared. By taking into
account that both $\nu $ and $z$ are large, we use the asymptotic (\ref%
{FnuLarge2}) for the functions $F_{\nu }^{D/2}(x)$ and $F_{\nu }^{D/2+1}(x)$
in (\ref{Tii}) and (\ref{T1D}). For the diagonal components, to the leading
order over $1/\alpha $ one gets $\langle T_{i}^{i}\rangle \approx \langle
T_{i}^{i}\rangle _{\mathrm{(M)}}^{(0)}$, where (no summation over $i$)%
\begin{eqnarray}
\langle T_{i}^{i}\rangle _{\mathrm{(M)}}^{(0)} &=&-\frac{\left( 4\pi \right)
^{-\frac{D}{2}}}{D\Gamma (D/2)}\mathbf{\,}\int_{m}^{\infty }d\lambda \,\frac{%
\left( \lambda ^{2}-m^{2}\right) ^{D/2}}{c_{1}(\lambda )c_{2}(\lambda
)e^{2a\lambda }-1}\,  \notag \\
&&\times \left[ 2{+}\frac{4D\left( \xi -\xi _{D}\right) w^{2}-m^{2}}{\lambda
^{2}-m^{2}}\sum_{j=1,2}{e}^{2{|{{x}^{1}-{a_{j}}}|}\lambda }c_{j}(\lambda )%
\right] ,  \label{TiiM0m}
\end{eqnarray}%
for the components $i\neq 1$ and%
\begin{equation}
\langle T_{1}^{1}\rangle _{\mathrm{(M)}}^{(0)}=\frac{2\left( 4\pi \right) ^{-%
\frac{D}{2}}}{\Gamma (D/2)}\mathbf{\,}\int_{m}^{\infty }d\lambda \,\frac{%
\lambda ^{2}\left( \lambda ^{2}-m^{2}\right) ^{D/2-1}}{c_{1}(\lambda
)c_{2}(\lambda )e^{2a\lambda }-1},  \label{T11M0m}
\end{equation}%
for the normal stress. These results coincide with the expressions given in
\cite{SahaRev} for the VEV\ of the energy-momentum tensor between two plates
in the Minkowski bulk. In the massless limit they are reduced to the
expressions in \cite{Rome02}. Note that the distribution of the normal
stress is uniform. For the off-diagonal component the leading order term in
the expansion over $1/\alpha $ is expressed as%
\begin{eqnarray}
\langle T_{D}^{1}\rangle &\approx &-\frac{2\left( 4\pi \right) ^{-\frac{D}{2}%
}}{\Gamma (D/2)\alpha }\mathbf{\,}\int_{m}^{\infty }d\lambda \,\frac{%
\sum_{j=1,2}\left( -1\right) ^{j}{e}^{2{|{{x}^{1}-{a_{j}}}|}\lambda
}c_{j}(\lambda )}{c_{1}(\lambda )c_{2}(\lambda )e^{2a\lambda }-1}  \notag \\
&&\times \lambda \left( \lambda ^{2}-m^{2}\right) ^{D/2-2}\left[ D\left( \xi
-\xi _{D}\right) \lambda ^{2}-\left( 2\xi -\frac{1}{4}\right) m^{2}\right] .
\label{T1DM0m}
\end{eqnarray}%
Of course, this component vanishes in the Minkowskian limit.

Now let us consider the special cases of Dirichlet and Neumann boundary
conditions. Similar to the discussion for the field squared, we expand the
function $1/(e^{2ax/z}-1)$ in (\ref{Tii}). The resulting integral over $x$
is presented in terms of the integral (\ref{Int2}) and its first and second
order derivatives with respect to $c$. In this way we can show that the VEVs
of the diagonal components of the energy-momentum tensor are presented as
(no summation over $i$)
\begin{equation}
\langle T_{i}^{i}\rangle =\langle T_{i}^{i}\rangle _{0}-\frac{\alpha ^{-1-D}%
}{2^{D+2\nu }\pi ^{\frac{D}{2}}}\mathbf{\,}\sum_{n=1}^{\infty }\left\{ \left[
\frac{E_{i}}{8}\partial _{c}^{2}h_{\nu }^{\frac{D}{2}}(c)+q_{\nu }^{(i)}(c)%
\right] _{c=\frac{na}{z}}{-}\frac{\delta _{\mathrm{J}}}{2}\sum_{j=1,2}\left.
q_{\nu }^{(i)}(c)\right\vert _{c=\frac{na-{|{{x}^{1}-{a_{j}}}|}}{z}}\right\}
.  \label{TiiDN}
\end{equation}%
Here we have defined the function%
\begin{equation}
q_{\nu }^{(i)}(c)=\left[ \left( w_{2}^{(i)}c^{2}+w^{(i)}\right) \partial
_{c}^{2}+w_{1}^{(i)}c\partial _{c}+w_{0}^{(i)}\right] h_{\nu }^{\frac{D}{2}%
}(c)+A_{i}h_{\nu }^{\frac{D}{2}+1}(c),  \label{qnui}
\end{equation}%
with the coefficients
\begin{equation}
(w_{2}^{(i)},w_{1}^{(i)},w_{0}^{(i)},w^{(i)})=\left( \xi -\frac{1}{4},D\xi -%
\frac{D+1}{4},-D\xi ,\left( \xi -\frac{1}{4}\right) \delta _{1}^{i}\right) ,
\label{wi}
\end{equation}%
for $i\neq D$ and%
\begin{equation}
(w_{2}^{(D)},w_{1}^{(D)},w_{0}^{(D)},w^{(D)})=\left( \frac{1}{4},D\xi +\frac{%
D+1}{4},D^{2}\xi -m^{2}\alpha ^{2},\xi \right) .  \label{wD}
\end{equation}%
For the off-diagonal component we get%
\begin{equation}
\langle T_{D}^{1}\rangle =\frac{\delta _{\mathrm{J}}\alpha ^{-1-D}}{%
2^{D+2\nu +3}\pi ^{\frac{D}{2}}}\sum_{n=1}^{\infty }\sum_{j=1,2}\left(
-1\right) ^{j}\left[ \left( 4\xi -1\right) c\partial _{c}-1\right] \partial
_{c}\left. h_{\nu }^{\frac{D}{2}}(c)\right\vert _{c=\frac{na-{|{{x}^{1}-{%
a_{j}}}|}}{z}}.  \label{T1DN}
\end{equation}%
This component has opposite signs for Dirichlet and Neumann boundary
conditions. Note that for the system of two scalars with Dirichlet and
Neaumann boundary conditions and with the same mass the total
energy-momentum tensor is diagonal and does not depend on the coordinate $%
x^{1}$.

The VEVs for a single brane with Dirichlet or Neumann boundary conditions
are obtained from (\ref{TiiDN}) in the limit when the location of the second
brane tends to infinity. For the diagonal components this gives (no
summation over $i$)%
\begin{equation}
\langle T_{i}^{i}\rangle =\langle T_{i}^{i}\rangle _{0}+\frac{\delta _{%
\mathrm{J}}\alpha ^{-1-D}}{2^{D+2\nu +1}\pi ^{\frac{D}{2}}}\mathbf{\,}q_{\nu
}^{(i)}\left( \frac{{|{{x}^{1}-{a_{j}}}|}}{z}\right) .  \label{TiiDN1}
\end{equation}%
In a similar way the expression for the off-diagonal component reads%
\begin{equation}
\langle T_{D}^{1}\rangle =\mathrm{sgn}(x^{1}-a_{j})\frac{\delta _{\mathrm{J}%
}\alpha ^{-1-D}}{2^{D+2\nu +3}\pi ^{\frac{D}{2}}}\left[ \left( 4\xi
-1\right) c\partial _{c}-1\right] \partial _{c}\left. h_{\nu }^{\frac{D}{2}%
}(c)\right\vert _{c={|{{x}^{1}-{a}}}_{j}|/z}.  \label{T1DN1}
\end{equation}%
Alternative representations for the VEVs (\ref{TiiDN1}) and (\ref{T1DN1}) in
terms of the function (\ref{fnu}) are provided in \cite{Beze15}.

The VEVs for the components of the energy-momentum tensor in the special
case of Dirichlet boundary condition on the brane $x^{1}=a_{1}$ and Neumann
condition on $x^{1}=a_{1}$ are obtained from (\ref{TiiDN}) and (\ref{T1DN})
by the replacements (\ref{DNrepl}).

Let us consider the behavior of the vacuum energy-momentum tensor near the
AdS boundary and horizon. Near the AdS boundary, assuming that $z\ll
|x^{1}-a_{j}|$ for $j=1,2$, the contributions of small $x$ dominate in (\ref%
{Tii}) and (\ref{T1D}). To the leading order, making the replacement $F_{\nu
}^{\frac{D}{2}}(x)\approx F_{\nu }^{\frac{D}{2}}(0)$, with $F_{\nu }^{\frac{D%
}{2}}(0)$ given by (\ref{Fmusmall}), we get (no summation over $i$)%
\begin{eqnarray}
\langle T_{i}^{i}\rangle &\approx &\langle T_{i}^{i}\rangle _{0}-\frac{%
B_{\nu }F_{\nu }^{\frac{D}{2}}(0)z^{D+2\nu }}{2^{D+2\nu }\pi ^{\frac{D-1}{2}%
}\alpha ^{D+1}}\left[ 2\nu -\left( D+2\nu \right) \delta _{i}^{D}\right]
\mathbf{\,}\int_{0}^{\infty }d\lambda \,\lambda ^{D+2\nu -1}\frac{2{+}%
\sum_{j=1,2}{e}^{2{|{{x}^{1}-{a_{j}}}|}\lambda }c_{j}(\lambda )}{%
c_{1}(\lambda )c_{2}(\lambda )e^{2a\lambda }-1},  \notag \\
\langle T_{D}^{1}\rangle &\approx &-\frac{2B_{\nu }F_{\nu }^{\frac{D}{2}%
}(0)z^{D+2\nu +1}}{2^{D+2\nu }\pi ^{\frac{D-1}{2}}\alpha ^{D+1}}\mathbf{\,}%
\int_{0}^{\infty }d\lambda \lambda ^{D+2\nu }\frac{\sum_{j=1,2}\left(
-1\right) ^{j}{e}^{2{|{{x}^{1}-{a_{j}}}|}\lambda }c_{j}(\lambda )}{%
c_{1}(\lambda )c_{2}(\lambda )e^{2a\lambda }-1},  \label{TiinearB}
\end{eqnarray}%
where we have defined%
\begin{equation}
B_{\nu }=\left( D+2\nu +1\right) \xi -\frac{D+2\nu }{4}.  \label{Bnu}
\end{equation}%
Under the conditions assumed, all the components tend to zero in the limit $%
z\rightarrow 0$. Note that the coefficient $B_{\nu }$ is negative for
minimally and conformally coupled fields.

For points tending to the horizon the coordinate $z$ is large. Assuming that
$z\gg a$ we use the asymptotic (\ref{Fmularge}) for the function $F_{\nu }^{%
\frac{D}{2}}(x)$. For the diagonal components this gives (no summation over $%
i$)%
\begin{equation}
\langle T_{i}^{i}\rangle \approx \langle T_{i}^{i}\rangle _{0}+(z/\alpha
)^{D+1}\langle T_{i}^{i}\rangle _{\mathrm{(M)}}^{(0)}|_{m=0},
\label{TiinearH}
\end{equation}%
with $\langle T_{i}^{i}\rangle _{\mathrm{(M)}}^{(0)}$ being the
corresponding VEV for two parallel plates in Minkowski spacetime given by (%
\ref{TiiM0m}) for a massive field. The leading term in the expansion of the
off-diagonal component is obtained from (\ref{T1DM0m}) taking $m=0$ and
multiplying by $(z/\alpha )^{D}$. For a non-conformally coupled field it
behaves like $(z/\alpha )^{D}$. For the conformal coupling the next term in
the expansion should be kept.

\begin{figure}[tbph]
\begin{center}
\begin{tabular}{cc}
\epsfig{figure=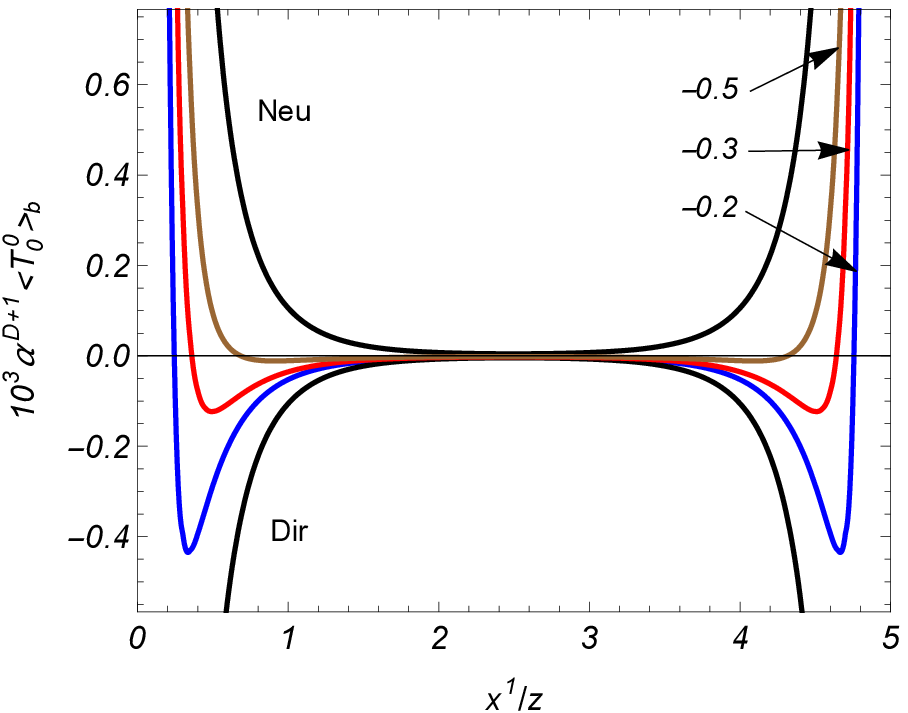,width=7.5cm,height=6cm} & \quad %
\epsfig{figure=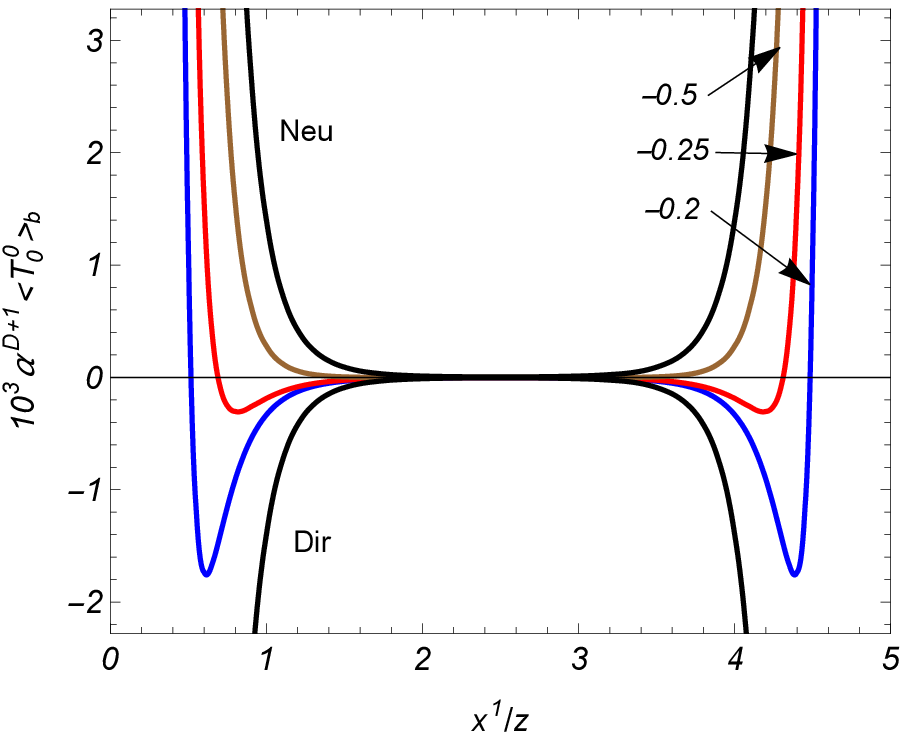,width=7.5cm,height=6cm}%
\end{tabular}%
\end{center}
\caption{The vacuum energy densities for $D=4$ conformally (left panel) and
minimally (right panel) coupled scalar fields induced by the branes in the
region $0<x^{1}/z<5$. The graphs are plotted in the cases of Dirichlet,
Neumann and Robin boundary conditions (with the values of $\protect\beta %
_{1}/z=\protect\beta _{1}/z$ given near the curves) for the locations of the
branes $a_{1}=0$, $a_{2}/z=5$ and for $m\protect\alpha =0.5$.}
\label{fig3}
\end{figure}

Figure \ref{fig3} presents the brane-induced energy density for conformally
(left panel) and minimally (right panel) coupled scalar fields in the region
between the branes versus the proper distance from the brane (in units of $%
\alpha $). The graphs are plotted for $a_{1}=0$, $a_{2}/z=5$, $m\alpha =0.5$
and for the same Robin boundary conditions on the branes ($\beta _{1}=\beta
_{2}$). The numbers near the graphs correspond to the values of the ratio $%
\beta _{1}/z$. We have also plotted the graphs for Dirichlet and Neumann
boundary conditions. In accordance with the asymptotic (\ref{Tnear}), for a
minimally coupled field and for non-Dirichlet boundary conditions the vacuum
energy density is positive near the branes. For Dirichlet boundary condition
the energy density is negative. The behavior of the energy density near the
center with respect to the branes depends on the Robin coefficients. For $%
\beta _{j}/z<0$ and sufficiently close to zero the brane-induced energy
density is negative near the center. With increasing value of $|\beta
_{j}|/z $, started from certain critical value $\beta _{j}^{(c)}$, that
depends on $a/z$, it becomes positive everywhere in the region between the
branes. For the values of the parameters corresponding to Figure \ref{fig3},
the critical values are given by $\beta _{j}^{(c)}/z\approx -1.12$ and $%
\beta _{j}^{(c)}/z\approx -0.69$ in the cases of conformal and minimal
couplings, respectively. The critical values $|\beta _{j}^{(c)}|/z$ are
increasing functions of $a/z$.

In this section we have considered the local densities induced by the
branes. They are well defined for points away from the branes and do not
contain renormalization ambiguities. The global quantities, such as the
total vacuum energy in the region between the branes (per unit surface of
the branes), are also of physical interest. However, because of the surface
divergences, it cannot be obtained by direct integration of the vacuum
energy density: an additional renormalization is required. This problem is
well-known from the theory of the Casimir effect for curved boundaries in
flat spacetime. It is worth mentioning that for general Robin boundary
conditions the vacuum energy obtained by the integration of the bulk energy
density, in general, does not coincide with the total vacuum energy
evaluated as the sum of the ground state energies for elementary
oscillators. As it has been discussed in \cite{Saha04} for general case of
the bulk and boundary geometries, the reason for that is the presence of
surface energy density located on constraining boundaries. For a scalar
field with general curvature coupling parameter the expression for the
surface energy-momentum tensor is obtained in \cite{Saha04} by using the
standard variational procedure. Similar to the case of the integrated bulk
energy, the corresponding VEV requires an additional renormalization. As an
example we can use the approach based on the generalized zeta function
approach. We plan to address these points in a separate publication.

\section{The Casimir forces}

\label{sec:Forces}

The $i$th component of the force acting on the surface element $dS$ of the
brane at $x^{1}=a_{j}$ is given by $-\langle T_{l}^{i}\rangle
_{x^{1}=a_{j}+0}n_{(+)j}^{l}dS$ in the region $x^{1}\geq a_{j}+0$ and by $%
-\langle T_{l}^{i}\rangle _{x^{1}=a_{j}-0}n_{(-)j}^{l}dS$ in the region $%
x^{1}\leq a_{j}-0$, where $n_{(\pm )j}^{l}=\pm \delta _{1}^{l}$. For the
resulting force we get
\begin{equation}
dF_{(j)}^{i}=\langle T_{1}^{i}\rangle |_{x^{1}=a_{j}+0}^{x^{1}=a_{j}-0}dS.
\label{dFj}
\end{equation}%
Due to the nonzero off-diagonal stress $\langle T_{1}^{D}\rangle $, in
addition to the normal component $dF_{(j)}^{1}$, this force has nonzero
component parallel to the brane (shear force), $dF_{(j)}^{D}$. First we will
consider the normal force.

\subsection{Normal force}

For the normal force acting on the brane at $x^{1}=a_{j}$ one has $%
dF_{(j)}^{1}=\langle T_{1}^{1}\rangle |_{z=a_{j}+0}^{z=a_{j}-0}dS$. For $%
\langle T_{1}^{1}\rangle $ we have the decomposition (\ref{Tii2}) in the
region between the branes and $\langle T_{1}^{1}\rangle =\langle
T_{1}^{1}\rangle _{j}$ in the remaining regions. The parts $\langle
T_{1}^{1}\rangle _{j}$ are the same on the left and right-hand sides of the
brane and they do not contribute to the net force. The nonzero contribution
comes from the part $\langle T_{1}^{1}\rangle -\langle T_{1}^{1}\rangle _{j}$
(given by the last term in (\ref{Tii2})) in the region between the branes.
Hence, for the vacuum effective pressure on the brane $x^{1}=a_{j}$, given
as $P_{j}=-\left( \langle T_{1}^{1}\rangle -\langle T_{1}^{1}\rangle
_{j}\right) _{x^{1}=a_{j}}$, one gets%
\begin{equation}
P_{j}=\frac{\alpha ^{-1-D}}{2^{D+2\nu }\pi ^{\frac{D-1}{2}}}\mathbf{\,}%
\int_{0}^{\infty }dx\,x\frac{-2+\left[ 2{+}c_{j}(x/z)+1/c_{j}(x/z)\right]
\hat{B}_{1}}{c_{1}(x/z)c_{2}(x/z)e^{2ax/z}-1}x^{D+2\nu }F_{\nu }^{\frac{D}{2}%
}(x).  \label{Pj}
\end{equation}%
The corresponding Casimir forces act on the sides $x^{1}=a_{1}+0$ and $%
x^{1}=a_{2}-0$. They are attractive for $P_{j}<0$ and repulsive for $P_{j}>0$%
. In the special cases of Dirichlet or Neumann boundary conditions the
Casimir forces are obtained directly from (\ref{TiiDN}) with $i=1$:%
\begin{equation}
P_{j}=-\frac{\alpha ^{-1-D}}{2^{D+2\nu +2}\pi ^{\frac{D}{2}}}\mathbf{\,}%
\sum_{n=1}^{\infty }\left[ \partial _{c}^{2}h_{\nu }^{\frac{D}{2}%
}(c)-4\left( 1{-}\delta _{\mathrm{J}}\right) q_{\nu }^{(1)}\left( c\right) %
\right] _{c=na/z}.  \label{PjDN}
\end{equation}%
For Dirichlet boundary condition on the brane $x^{1}=a_{1}$ and Neumann
condition for $x^{1}=a_{2}$ the corresponding formula is obtained from (\ref%
{PjDN}) by the replacement (\ref{DNrepl}).

The expression for the Casimir normal force in the Minkowskian limit
directly follows from (\ref{T11M0m}). The corresponding effective pressure
is expressed as $P_{j}^{\mathrm{(M)}}=-\langle T_{1}^{1}\rangle _{\mathrm{(M)%
}}^{(0)}$. Note that for the Minkowskian bulk the forces acting on separate
plates coincide regardless of the values of the Robin coefficients. As seen
from (\ref{Pj}), in general this is not the case for the AdS geometry.

For small separations between the branes, $a\ll z$, the dominant
contribution to the integral in (\ref{Pj}) comes from the region with large $%
x$ and we use the asymptotic (\ref{Fmularge}) for the function $F_{\nu
}^{\mu }(x)$. The leading term in the expansion of the force comes from the
part with -2 in the numerator of the integrand in (\ref{Pj}) and we get%
\begin{equation}
P_{j}\approx -\frac{2\left( z/\alpha \right) ^{D+1}}{(4\pi )^{\frac{D}{2}%
}\Gamma \left( \frac{D}{2}\right) }\int_{0}^{\infty }d\lambda \,\frac{%
\lambda ^{D}}{c_{1}(\lambda )c_{2}(\lambda )e^{2a\lambda }-1}.
\label{pjsmall}
\end{equation}%
If additionally one has $a\ll |\beta _{l}|$, $l=1,2$, we note that the
integral in (\ref{pjsmall}) is dominated by the contribution from the region
$\lambda \lesssim 1/a$ and in that region $c_{1}(\lambda )c_{2}(\lambda
)\approx 1$. The estimate (\ref{pjsmall}) is further simplified as%
\begin{equation}
P_{j}\approx -\frac{D\zeta (D+1)}{\left( 2\sqrt{\pi }\alpha a/z\right) ^{D+1}%
}\Gamma \left( \frac{D+1}{2}\right) ,  \label{pjsmall1}
\end{equation}%
where $\zeta (x)$ is the Riemann zeta function. For Dirichlet boundary
conditions on both the branes $c_{1}(\lambda )c_{2}(\lambda )=1$ and we get
the same leading term. For Dirichlet boundary condition on one brane and
non-Dirichlet condition on the other, with the modulus of the Robin
coefficient much larger than $a$, we have $c_{1}(\lambda )c_{2}(\lambda
)\approx -1$. In this case (\ref{pjsmall}) is reduced to%
\begin{equation}
P_{j}\approx \frac{D\zeta (D+1)}{\left( 2\sqrt{\pi }\alpha a/z\right) ^{D+1}}%
\left( 1-\frac{1}{2^{D}}\right) \Gamma \left( \frac{D+1}{2}\right) .
\label{pjsmall2}
\end{equation}%
The approximations (\ref{pjsmall1}) and (\ref{pjsmall2}) are obtained from
the corresponding asymptotics for Robin plates in the Minkowski spacetime
replacing the separation between the plates by the proper separation $\alpha
a/z$ in the AdS bulk. The asymptotics show that for small separations
between the branes ($a\ll z$ and $a\ll |\beta _{l}|$ for non-Dirichlet
boundary conditions) the Casimir normal forces are repulsive for Dirichlet
boundary condition on one brane and non-Dirichlet condition on the other
(formula (\ref{pjsmall2})). In the remaining cases the forces are
attractive. In the asymptotic region under consideration with the proper
separation much smaller than the curvature radius, the effects of gravity on
the Casimir forces are small and the results are similar to those for the
Minkowski bulk.

We expect that the influence of the gravity will be essential for proper
separations larger than the AdS curvature radius. In the limit $a/z\gg 1$
the integral in (\ref{Pj}) is dominated by the contribution from the region
with small $x$. Expanding the function $F_{\nu }^{D/2}(x)$ in (\ref{Pj}) one
finds%
\begin{equation}
P_{j}\approx \frac{2\pi ^{\frac{1-D}{2}}\alpha ^{-1-D}\left( z/2\right)
^{D+2\nu }}{\Gamma \left( \frac{D+1}{2}+\nu \right) \Gamma \left( 1+\nu
\right) }\mathbf{\,}\int_{0}^{\infty }d\lambda \,\lambda ^{D+2\nu -1}\frac{%
\nu B_{\nu }\left[ 2{+}c_{j}(\lambda )+1/c_{j}(\lambda )\right] -\lambda
^{2}z^{2}}{c_{1}(\lambda )c_{2}(\lambda )e^{2a\lambda }-1},  \label{Pjlarge}
\end{equation}%
where $B_{\nu }$ is defined by (\ref{Bnu}). Under additional conditions $%
a\gg |\beta _{l}|$, $l=1,2$ (non-Neumann boundary conditions on both the
branes), we further expand the integrand over the small ratio $|\beta _{l}|/a
$ with the result%
\begin{equation}
P_{j}\approx -\frac{2\left( D+2\nu +1\right) \left( 4\nu B_{\nu }\beta
_{j}^{2}/z^{2}+1\right) }{\pi ^{\frac{D}{2}}\Gamma \left( 1+\nu \right)
\alpha ^{D+1}\left( 2a/z\right) ^{D+2\nu +2}}\zeta \left( D+2\nu +2\right)
\Gamma \left( \frac{D}{2}+\nu +1\right) ,  \label{Pjlarge1}
\end{equation}%
For $a\gg |\beta _{j}|$ and for Neumann boundary condition on the second
brane ($c_{j^{\prime }}(\lambda )=1$, as before, $j^{\prime }=1$ for $j=2$
and $j^{\prime }=2$ for $j=1$) from (\ref{Pjlarge}) we get
\begin{eqnarray}
P_{j} &\approx &\frac{2\left( D+2\nu +1\right) \left( 4\nu B_{\nu }\beta
_{j}^{2}/z^{2}+1\right) }{\pi ^{\frac{D}{2}}\Gamma \left( 1+\nu \right)
\alpha ^{D+1}\left( 2a/z\right) ^{D+2\nu +2}}\mathbf{\,}  \notag \\
&&\times \left( 1-\frac{1}{2^{D+2\nu +1}}\right) \zeta \left( D+2\nu
+2\right) \Gamma \left( \frac{D}{2}+\nu +1\right) .  \label{Pjlarge2}
\end{eqnarray}%
The forces corresponding to (\ref{Pjlarge1}) and (\ref{Pjlarge2}) have
opposite signs. As it has been already mentioned before, the coefficient $%
B_{\nu }$ is negative for minimally and conformally coupled fields. Then,
from (\ref{Pjlarge1}) we see that, depending on the boundary conditions, the
Casimir forces can be either attractive or repulsive at large distances. The
sign of the forces is determined by the factor $4\nu B_{\nu }\beta
_{j}^{2}/z^{2}+1$. This factor is positive near the horizon and is negative
near the AdS boundary if $B_{\nu }<0$. This shows that, for given values of
the parameters, the vacuum pressure changes the sign as a function of $z$.

For Neumann boundary condition on the brane at $x^{1}=a_{j}$ and at large
separations, to the leading order, we can ignore the term $\lambda ^{2}z^{2}$
in (\ref{Pjlarge}). For non-Neumann boundary condition on the second brane,
assuming $a\gg |\beta _{j^{\prime }}|$, this gives%
\begin{equation}
P_{j}\approx -\frac{4\nu B_{\nu }\left( 1-2^{1-D-2\nu }\right) \zeta (D+2\nu
)}{\pi ^{\frac{D}{2}}\Gamma \left( 1+\nu \right) \alpha ^{D+1}\left(
2a/z\right) ^{D+2\nu }}\Gamma \left( \frac{D}{2}+\nu \right) .
\label{PjNlarge}
\end{equation}
By taking into account that $B_{\nu }<0$ for minimal and conformal
couplings, this result shows that for Neumann boundary condition on the
brane $x^{1}=a_{j}$ and for non-Neumann condition on the second brane the
force is repulsive at large separations. For Neumann boundary condition on
both the branes the leading term is expressed as%
\begin{equation}
P_{j}\approx \frac{4\nu B_{\nu }\zeta \left( D+2\nu \right) \Gamma \left(
D/2+\nu \right) }{\pi ^{\frac{D}{2}}\Gamma \left( 1+\nu \right) \alpha
^{D+1}\left( 2a/z\right) ^{D+2\nu }},  \label{PjNlarge1}
\end{equation}%
and the force is attractive for $B_{\nu }<0$. The decay of the normal force
at large proper separations between the branes is power-law for both
massless and massive cases. In the Minkowski bulk and for massive fields the
corresponding suppression is exponential. The leading term is found from (%
\ref{T11M0m}), $P_{j}^{\mathrm{(M)}}\propto a^{-D/2}e^{-2ma}$.

As seen from the analysis given above, for the brane with Neumann boundary
condition the Casimir force on that brane decays at large separations like $%
(z/a)^{D+2\nu }$ regardless the boundary condition on the second brane
(except the special case with $B_{\nu }=0$). For non-Neumann boundary
conditions on the brane at $x^{1}=a_{j}$ and for $a\gg |\beta _{j}|$ the
corresponding force behaves as $(z/a)^{D+2\nu +2}$ and the suppression is
stronger. As an example let us consider the case of Dirichlet boundary
condition for $x^{1}=a_{1}$ and Neumann condition for $x^{1}=a_{2}$. At
large separations the Casimir pressure on the brane $x^{1}=a_{1}$ is given
by (\ref{Pjlarge2}) with $j=1$ and $\beta _{j}=0$. It corresponds to a
repulsive force. The leading term for the Casimir force on the brane $%
x^{1}=a_{2}$ is obtained from (\ref{PjNlarge}) with $j=2$. The force is
repulsive for $B_{\nu }<0$ and attractive for $B_{\nu }>0$. This shows that,
in principle, we can have a situation when the force has an attractive
nature for one brane and repulsive nature for another.

\begin{figure}[tbph]
\begin{center}
\epsfig{figure=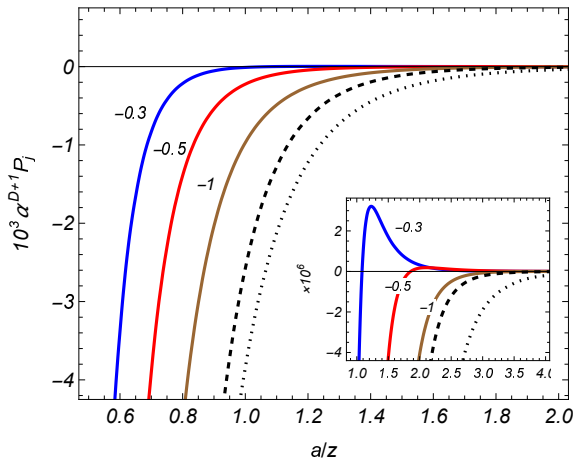,width=7.5cm,height=6cm}
\end{center}
\caption{The Casimir normal force for $D=4$ minimally coupled field with $m%
\protect\alpha =0.5$ as a function of the interbrane separation. The graphs
are plotted for Dirichlet and Neumann boundary conditions (dashed and dotted
curves), and for Robin boundary conditions with the coefficients $\protect%
\beta _{1}/z=\protect\beta _{2}/z$ given near the corresponding graphs. }
\label{fig4}
\end{figure}

In Figure \ref{fig4} we have displayed the normal Casimir force versus the
proper separation between the branes, in units of the AdS curvature radius,
for $D=4$ minimally coupled scalar field. The same boundary conditions are
imposed on the branes. The numbers near the curves are the values for $\beta
_{1}/z=$ $\beta _{2}/z$. The dashed and dotted curves correspond to
Dirichlet and Neumann boundary conditions, respectively. The graphs are
plotted for $m\alpha =0.5$. The presented graphs demonstrate the feature
already seen from asymptotic analysis: the forces attractive at small
separations may become repulsive for larger distances.

\begin{figure}[tbph]
\begin{center}
\begin{tabular}{cc}
\epsfig{figure=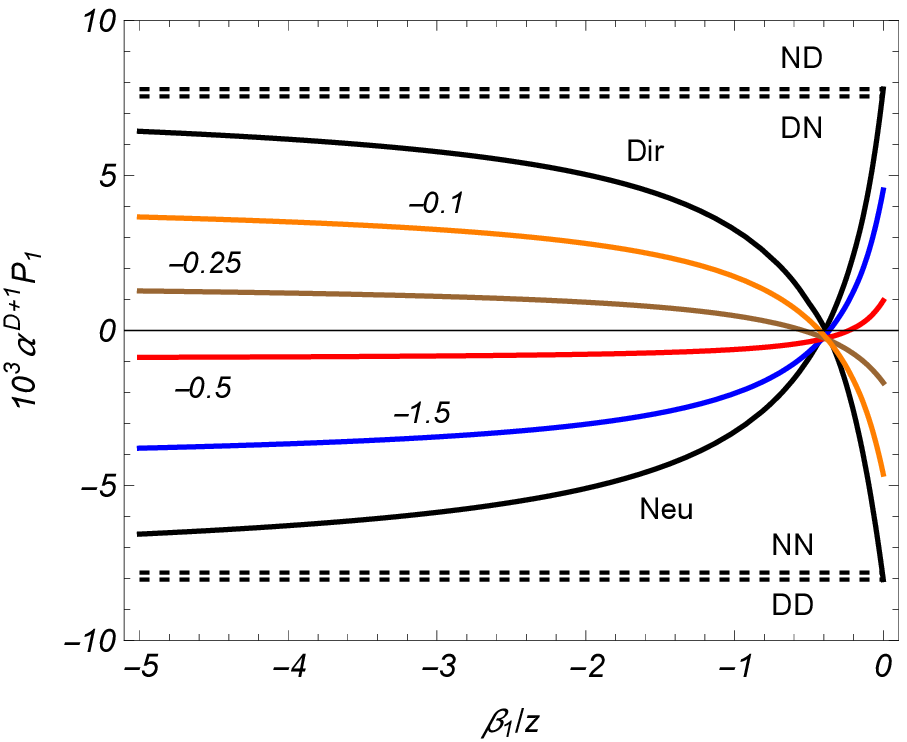,width=7.5cm,height=6.5cm} & \quad %
\epsfig{figure=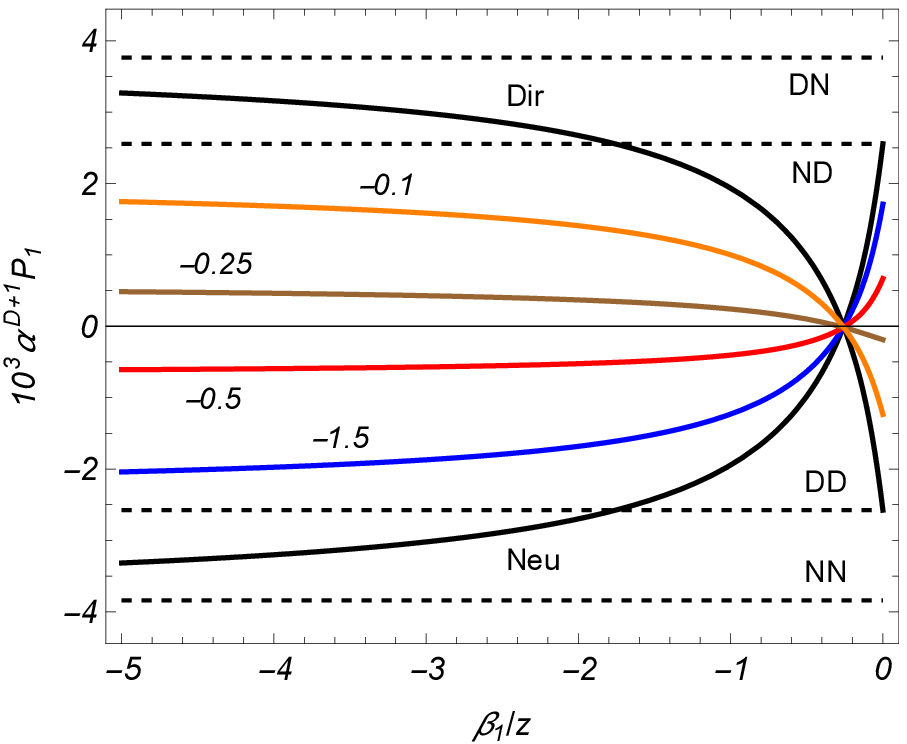,width=7.5cm,height=6.5cm}%
\end{tabular}%
\end{center}
\caption{The Casimir normal force per unit surface of the brane $x^{1}=a_{1}$
as a function of the Robin coefficient in the boundary condition on that
brane for $D=4$ conformally (left panel) and minimally (right panel) coupled
fields. The graphs are plotted for $m\protect\alpha =0.5$, $a/z=1$, and for
different boundary conditions on the second brane (see the text). }
\label{fig5}
\end{figure}

Figure \ref{fig5} presents the dependence of the Casimir normal force acting
on the brane at $x^{1}=a_{1}$, given by (\ref{Pj}) with $j=1$, on the
coefficient in the Robin boundary condition on that brane. The left and
right panels correspond to $D=4$ conformally and minimally coupled fields
with $m\alpha =0.5$. For the proper separation between the branes we have
taken $a/z=1$. The graphs are plotted for different boundary conditions on
the second brane: Dirichlet and Neumann conditions (Dir and Neu,
respectively), Robin boundary conditions with $\beta _{2}/z$ given near the
curves. The dashed lines correspond to Dirichlet and Neumann conditions on
both the branes (DD and NN), Dirichlet (Neumann) condition at $x^{1}=a_{1}$
and Neumann (Dirichlet) condition at $x^{1}=a_{2}$, indicated as DN (ND).
The graphs show that depending on the coefficient in the Robin boundary
conditions the force can be either attractive or repulsive.

\subsection{Shear force}

As it has been emphasized above, in the problem at hand in addition to the
normal Casimir force one has a nonzero shear force along the $z$-direction, $%
dF_{(j)}^{D}=f_{(j)}dS$, where $f_{(j)}$ is the shear force per unit surface
of the plate at $z=z_{j}$. The latter is given by $f_{(j)}=\langle
T_{1}^{D}\rangle |_{x^{1}=a_{j}+0}^{x^{1}=a_{j}-0}$. In accordance with the
decomposition (\ref{T1D2}), the shear force contains two contributions. The
first part comes from the term $\langle T_{1}^{D}\rangle _{j}$ and
corresponds to the force acting on the brane at $x^{1}=a_{j}$ when the
second brane is absent. We will call this part the self-acting shear force
and will denote by $f_{j}^{\mathrm{(s)}}$. Those forces acting on the sides $%
x^{1}=a_{j}-0$ and $x^{1}=a_{j}+0$ coincide and we get%
\begin{equation}
f_{j}^{\mathrm{(s)}}=\langle T_{1}^{D}\rangle
_{j}|_{x^{1}=a_{j}+0}^{x^{1}=a_{j}-0}=\frac{4\alpha ^{-1-D}}{2^{D+2\nu }\pi
^{\frac{D-1}{2}}}\mathbf{\,}\int_{0}^{\infty }dx\,\frac{{1}}{c_{j}(x/z)}%
\left[ \left( \xi -\frac{1}{4}\right) x\partial _{x}+\xi \right] x^{D+2\nu
}F_{\nu }^{\frac{D}{2}}(x).  \label{fjs}
\end{equation}%
By using the asymptotic (\ref{Fmularge}), we see that for nonconformally
coupled fields and for large $x$ the integrand in (\ref{fjs}) behaves like $%
x^{D-1}$ and the integral is divergent in the upper limit. For the conformal
coupling the next to the leading term should be kept and the integral is
still divergent. Of course, the divergence comes from the surface
divergences in the single brane contributions to the VEVs. The
renormalization of the divergence in the self-action shear force can be
considered in the same line as that for the total and surface Casimir
energies and will be discussed elsewhere. Here we will be focused on the
contribution to the shear force that is induced by the second brane. This
part acts on the sides $x^{1}=a_{1}+0$ and $x^{1}=a_{2}-0$ and is determined
from the last term in (\ref{T1D2}). Denoting it by $f_{j}^{\mathrm{(int)}}$,
we get
\begin{equation}
f_{j}^{\mathrm{(int)}}=-\frac{2\alpha ^{-1-D}}{2^{D+2\nu }\pi ^{\frac{D-1}{2}%
}}\mathbf{\,}\int_{0}^{\infty }dx\frac{c_{j}(x/z)-1/c_{j}(x/z)}{%
c_{1}(x/z)c_{2}(x/z)e^{2ax/z}-1}\left[ \left( \xi -\frac{1}{4}\right)
x\partial _{x}+\xi \right] x^{D+2\nu }F_{\nu }^{\frac{D}{2}}(x).
\label{fjint}
\end{equation}%
This part acting on the brane at $x^{1}=a_{j}$ vanishes for Dirichlet and
Neumann boundary conditions on that brane regardless of boundary conditions
on the second brane. The shear force is directed toward the horizon for $%
f_{j}^{\mathrm{(int)}}>0$ and toward the AdS boundary for $f_{j}^{\mathrm{%
(int)}}<0$.

The asymptotic behavior of the shear force is found in a way similar to that
for the normal force. At small proper separations compared with the
curvature radius, $a/z\ll 1$, one gets%
\begin{equation}
f_{j}^{\mathrm{(int)}}\approx -\frac{2D\left( \xi -\xi _{D}\right) z^{D}}{%
2^{D}\pi ^{\frac{D}{2}}\Gamma \left( \frac{D}{2}\right) \alpha ^{D+1}}%
\int_{0}^{\infty }d\lambda \,\lambda ^{D-1}\frac{c_{j}(\lambda
)-1/c_{j}(\lambda )}{c_{1}(\lambda )c_{2}(\lambda )e^{2a\lambda }-1}.
\label{fjsmall}
\end{equation}%
For a conformally coupled field the leading term vanishes and the next term
in the expansion should be kept. If additionally $\left\vert \beta
_{l}\right\vert \gg a$, $l=1,2$ (the condition with $l=j^{\prime }$ is
required only for non-Dirichlet boundary conditions on the brane at $%
x^{1}=a_{j^{\prime }}$), the further expansion gives
\begin{equation}
f_{j}^{\mathrm{(int)}}\approx \frac{4D\left( \xi -\xi _{D}\right) \zeta (D-1)%
}{\pi ^{\frac{D+1}{2}}\alpha ^{D+1}\left( 2a/z\right) ^{D}b_{j}}\Gamma
\left( \frac{D-1}{2}\right) \left( 2^{2-D}-1\right) ^{\delta _{0b_{j^{\prime
}}}},  \label{fjsmall2}
\end{equation}%
where the last factor is present only for Dirichlet boundary condition at $%
x^{1}=a_{j^{\prime }}$. Note that under the specified conditions one has $%
|b_{j}|\gg 1$. As seen, at small separations, the shear component of the
force has opposite signs for Dirichlet and non-Dirichlet boundary conditions
on the second brane. For a minimally coupled field with $b_{j}<0$ and for
small separations the shear force acting on the brane at $x^{1}=a_{j}$ is
directed toward the AdS horizon for non-Dirichlet boundary conditions on the
second brane and toward the AdS boundary for Dirichlet condition.

At large proper separations, $a/z\gg 1$, the interaction force is
approximated by
\begin{equation}
f_{j}^{\mathrm{(int)}}\approx -\frac{2\pi ^{\frac{1-D}{2}}B_{\nu }\alpha
^{-1-D}\left( z/a\right) ^{D+2\nu +1}}{2^{D+2\nu }\Gamma \left( \nu
+1\right) \Gamma \left( \frac{D+1}{2}+\nu \right) }\mathbf{\,}%
\int_{0}^{\infty }dx\,\frac{c_{j}(x/a)-1/c_{j}(x/a)}{%
c_{1}(x/a)c_{2}(x/a)e^{2x}-1}x^{D+2\nu }.  \label{fjlarge}
\end{equation}%
This estimate is further simplified under the condition $\left\vert \beta
_{l}\right\vert \ll a$, $l=1,2$ (the condition for $l=j^{\prime }$ is
required only for non-Neumann boundary conditions at $x^{1}=a_{j^{\prime }})$%
:%
\begin{equation}
f_{j}^{\mathrm{(int)}}\approx -\frac{4b_{j}B_{\nu }\left( D+2\nu +1\right)
\zeta \left( D+2\nu +2\right) }{\pi ^{\frac{D}{2}}\Gamma \left( \nu
+1\right) \alpha ^{D+1}\left( 2a/z\right) ^{D+2\nu +1}}\Gamma \left( \frac{D%
}{2}+\nu +1\right) \left( \frac{1}{2^{D+2\nu +1}}-1\right) ^{\delta _{\infty
b_{j^{\prime }}}},  \label{fjlarge2}
\end{equation}%
where $B_{\nu }$ is defined by (\ref{Bnu}) and $|b_{j}|\ll 1$. The force (%
\ref{fjlarge2}) has opposite signs for Neumann and non-Neumann boundary
conditions on the brane $x^{1}=a_{j^{\prime }}$. For conformally and
minimally coupled fields one has $B_{\nu }<0$. In those cases, for $b_{j}<0$
and at large separations between the branes the shear force acting on the
brane $x^{1}=a_{j}$ is directed toward the AdS horizon for Neumann boundary
condition on the second brane and toward the AdS boundary for non-Neumann
conditions.

\begin{figure}[tbph]
\begin{center}
\begin{tabular}{cc}
\epsfig{figure=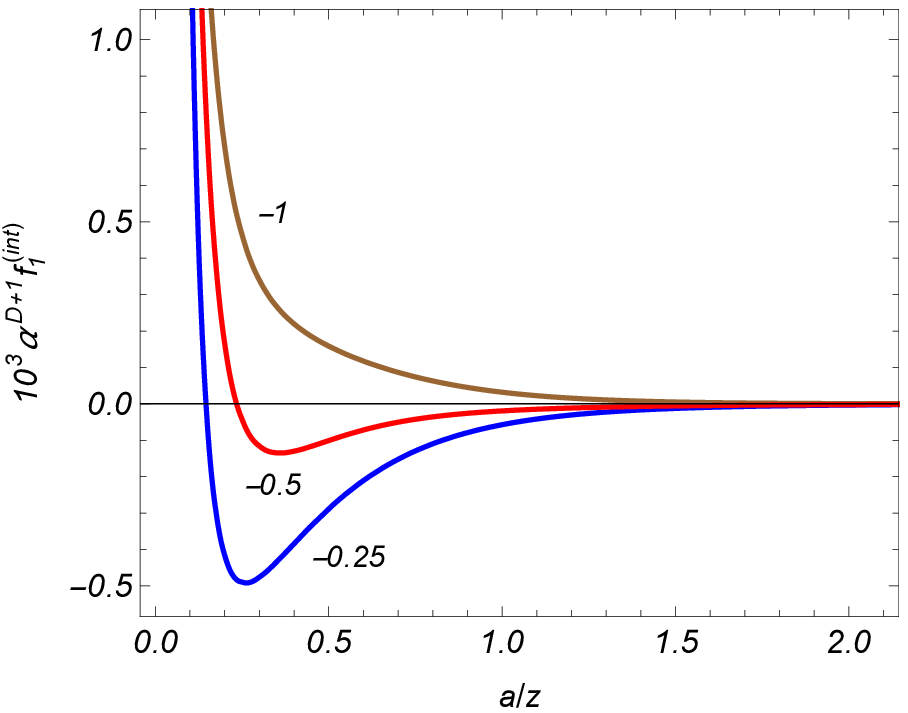,width=7.5cm,height=6cm} & \quad %
\epsfig{figure=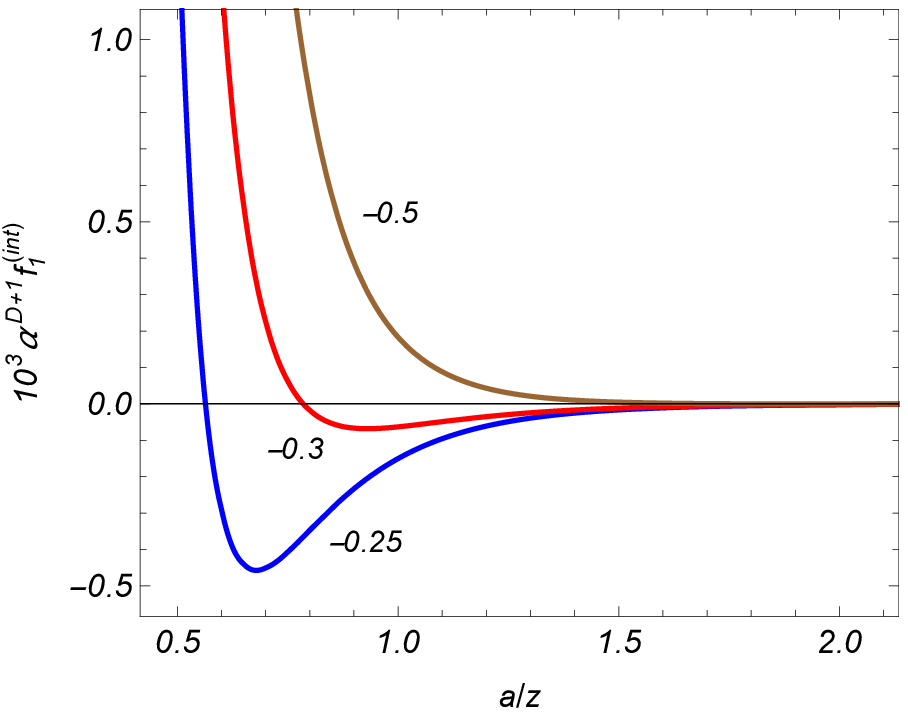,width=7.5cm,height=6cm}%
\end{tabular}%
\end{center}
\caption{The interaction contribution to the shear force per unit surface of
the brane $x^{1}=a_{1}$ versus the interbrane separation for different
values of $\protect\beta _{1}/z=\protect\beta _{2}/z$ (the numbers given
near the curves). The graphs are plotted for $D=4$ conformally (left panel)
and minimally (right panel) coupled fields with $m\protect\alpha =0.5$. }
\label{fig6}
\end{figure}

The interaction part of the shear force acting on the brane $x^{1}=a_{1}$
versus the distance between the branes is depicted in Figure \ref{fig6} for $%
D=4$ conformally and minimally coupled field (left and right panels,
respectively). The graphs are plotted for $m\alpha =0.5$ and for different
values of the ratio $\beta _{1}/z=\beta _{2}/z$ (the numbers near the
curves). In both cases the shear force is directed toward the horizon at
small separations between the branes and toward the AdS boundary at large
separations. For a minimally coupled field this is in agreement with the
asymptotic analysis presented above.

\begin{figure}[tbph]
\begin{center}
\begin{tabular}{cc}
\epsfig{figure=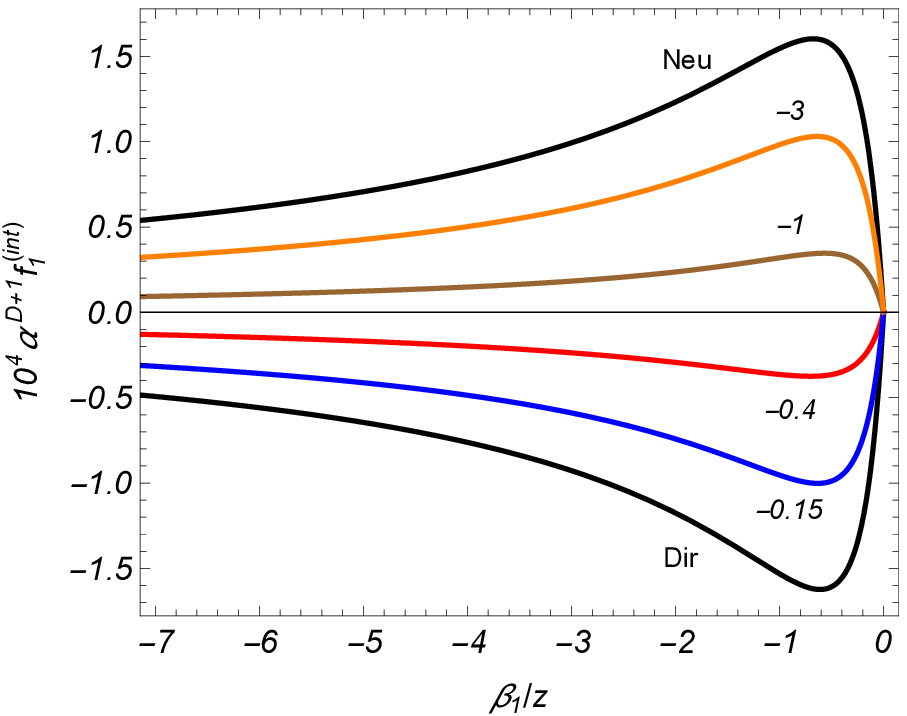,width=7.5cm,height=6.5cm} & \quad %
\epsfig{figure=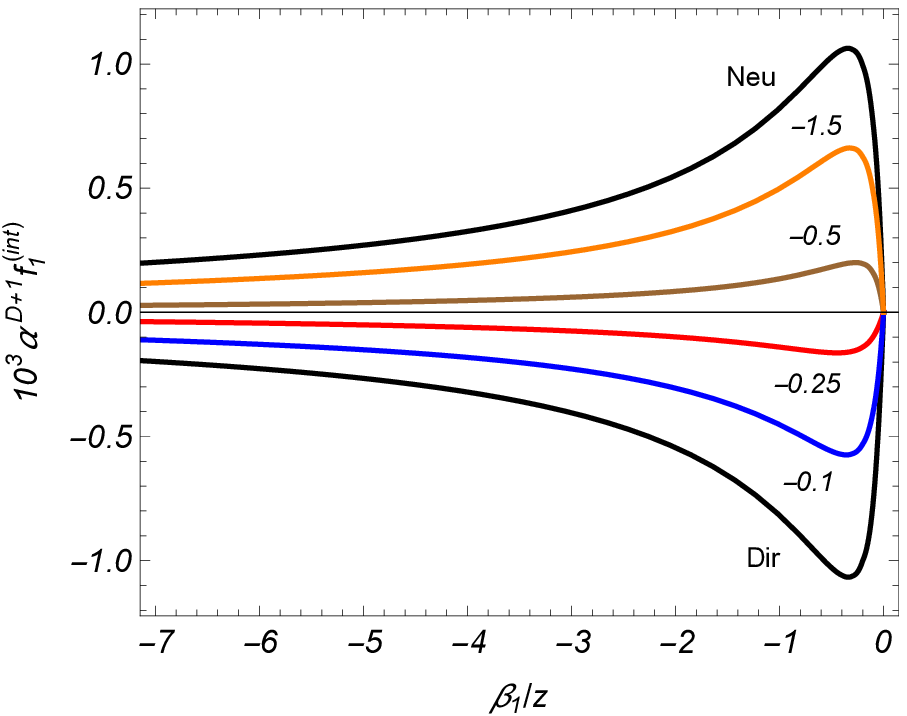,width=7.5cm,height=6.5cm}%
\end{tabular}%
\end{center}
\caption{The interaction shear force acting on the brane $x^{1}=a_{1}$
versus the corresponding Robin coefficient for $D=4$ conformally (left
panel) and minimally (right panel) coupled fields. The numbers near the
curves are the values of the ratio $\protect\beta _{2}/z$ and the graphs are
plotted for $m\protect\alpha =0.5$, $a/z=1$. }
\label{fig7}
\end{figure}

The interaction part of the shear force per unit surface of the brane $%
x^{1}=a_{1}$ is plotted in Figure \ref{fig7} as a function of the ratio $%
\beta _{1}/z$ for different boundary conditions on the second brane
(Dirichlet, Neumann and Robin conditions with the values for $\beta _{2}/z$
presented near the curves). The left and right panels correspond to
conformally and minimally coupled fields in (4+1)-dimensional AdS spacetime.
The graphs are plotted for $m\alpha =0.5$ and $a/z=1$.

\section{Conclusion}

\label{sec:Conc}

In this paper we have investigated the influence of two parallel branes,
orthogonal to the AdS boundary, on the local properties of the scalar vacuum
in background of $(D+1)$-dimensional AdS spacetime. Robin boundary
conditions are imposed, in general, with different coefficients on separate
branes. We consider a free field theory and the properties of the vacuum are
completely determined by the two-point functions. As a two-point function,
the positive frequency Wightman function is chosen. The local VEVs are
obtained in the coincidence limit of the arguments of that function and its
derivatives. For the evaluation of the Wightman function the direct
summation over the complete set of scalar modes is used. In the region
between the branes the mode functions are given by (\ref{mf}) with the
function $\alpha _{j}(\lambda )$ defined as (\ref{alfj}). The eigenvalues of
the quantum number $\lambda $ are discretized by the boundary conditions and
they are expressed in terms of the roots of equation (\ref{meq}). The
geometry of the subspace $y=\mathrm{const}$, parallel to the AdS boundary,
is Minkowskian and the eigenvalue equation coincides with that in the
Casimir problem for two Robin plates in flat spacetime. For general Robin
boundary conditions the eigenvalues of $\lambda $ are given implicitly and
for the summation of the corresponding series in the mode sum of the
Wightman function we have employed the Abel-Plana-type formula (\ref{APf}).
This has two advantages: (i) an integral representation is provided for
which the explicit knowledge of the eigenvalues is not required and (ii) the
parts corresponding to the brane-free and single brane geometries are
explicitly extracted. In particular, on the basis of (ii), the
renormalization of local VEVs for points outside the branes is reduced to
the one in the brane-free problem.

As a local characteristic of the vacuum state we have considered the mean
field squared. Based on the Wightman function decomposition, the VEV is
presented in two equivalent forms, (\ref{ph23}) and (\ref{ph24}). In the
second one the contribution corresponding to the problem with a single brane
is separated. For special cases of Dirichlet and Neumann boundary conditions
the VEVs are further simplified to (\ref{ph2DN}). An alternative
representation for those cases is given by (\ref{ph2DN2}). For a conformally
coupled massless scalar field the problem under consideration is conformally
related to the problem with parallel Robin plates in the Minkowski spacetime
orthogonally intersected by a Dirichlet plate, the latter being the
conformal image of the AdS boundary. The Dirichlet boundary condition on the
conformal image is related to the condition for the field modes (\ref{mf})
imposed on the AdS boundary. In the Minkowskian limit we recover the result
for a massive scalar field in the geometry of two parallel plates,
previously considered in \cite{Rome02} and \cite{SahaRev} for massless and
massive fields, respectively. For points near the branes and not too close
to the AdS boundary the dominant contribution to the VEV comes from quantum
fluctuations with wavelengths smaller than the curvature radius and the
influence of the gravity is weak. The leading term in the expansion over the
distance from the brane coincides with that for a plate in the Minkowski
bulk with the distance from the plate replaced by the proper distance in the
AdS bulk. The brane-induced contribution vanishes on the AdS boundary. For
points not too close to the branes the corresponding asymptotic is given by (%
\ref{ph2nearAdSb}). In the opposite near-horizon limit, for fixed value of
the coordinate distance $a$, the proper separation between the branes is
small compared to the curvature radius and the brane-induced VEV is well
approximated by the Minkowskian expression (see (\ref{ph2nearH})). Depending
on the boundary conditions, the mean field squared, as a function of the
distance from the brane, may change the sign.

The vacuum energy density and stresses in the region between the branes have
been discussed in Section \ref{sec:EMT}. The diagonal components of the
vacuum energy-momentum tensor are given by the formula (\ref{Tii}). The only
nonzero off-diagonal component corresponds to the stress $\langle
T_{D}^{1}\rangle $, expressed as (\ref{T1D}). The generation of this
component is a pure brane-induced effect and gives rise to a shear force
acting on the branes. As expected, the brane-induced contribution obeys the
trace relation (\ref{Trace}) and the covariant conservation equation. Single
brane contributions in the components of the vacuum energy-momentum tensor
are explicitly separated in the representations (\ref{Tii2}) and (\ref{T1D2}%
). In the Minkowskian limit we recover the results of Refs. \cite{Rome02}
and \cite{SahaRev} for massless and massive scalar fields. In the special
case of a conformally coupled massless field the brane-induced part has a
conformal connection with the corresponding vacuum energy-momentum tensor
for two parallel plates with Robin boundary conditions intersected by the
third plate with Dirichlet boundary condition. The respective VEVs are given
by (\ref{TiiM}) and (\ref{T1DM}). For special cases of Dirichlet and Neumann
boundary conditions equivalent representations are given by formulae (\ref%
{TiiDN}) and (\ref{T1DN}). Near the branes and near the horizon, for fixed
value of the separation $a$, the effects of the gravity on the brane-induced
VEVs of the components $\langle T_{i}^{i}\rangle $ with $i\neq 1$ are weak
and the leading terms in the corresponding expansions coincide with those
for the Minkowski bulk. The brane-induced contributions in the diagonal
components vanish on the AdS boundary like $z^{D+2\nu }$. The decay for the
off-diagonal component is stronger, as $z^{D+2\nu +1}$. The numerical
investigation for the distribution of the vacuum energy density is presented
for the case when the boundary conditions imposed on separate branes are the
same. The brane-induced vacuum energy density in the region between the
branes is negative for Dirichlet boundary conditions and positive for
Neumann conditions. For Robin conditions there is a critical value of the
coefficient $\beta _{j}=\beta _{j}^{(c)}<0$ that separates two qualitatively
different distributions. For $\beta _{j}<\beta _{j}^{(c)}$ the behavior of
the energy density is of Neuman-type: the energy density is positive
everywhere in the region between the branes. In the range $\beta
_{j}^{(c)}<\beta _{j}<0$ the energy density is positive near the branes and
negative near the center with respect to the brane locations. This type of
beahvior is depicted in Figure \ref{fig3}.

The Casimir forces acting on the branes have two components. The first one
corresponds to the normal force which is considered in the literature for
different bulk and boundary geometries. Interpreted in terms of the vacuum
pressure on the brane at $x^{1}=a_{j}$, it is given by the expression (\ref%
{Pj}) or by an alternative representation (\ref{PjDN}) for Dirichlet and
Neumann boundary conditions. Unlike the problem in the Minkowskian bulk, the
forces for Dirichlet and Neumann boundary conditions are different. Another
difference is that the forces acting on separate branes differ if the
coefficients in the Robin boundary conditions on them are different.
Depending on the boundary conditions and on the separation between the
branes the normal forces can be either attractive or repulsive. At small
separations the effects of background curvature are weak and the force is
well approximated by the corresponding result for the Minkowski bulk. They
are repulsive for Dirichlet boundary condition on one brane and
non-Dirichlet condition on the other and attractive in the remaining cases.
The influence of gravity is essential for proper separations larger than the
AdS curvature radius. The decay of forces at large separations is power-law
for both cases of massless and massive fields. For massive fields this
results is in contrast to the exponential decay in the Minkowski bulk. The
Casimir normal force acting on the brane decays at large separations like $%
(z/a)^{D+2\nu }$ for Neumann boundary condition on that brane and behaves as
$(z/a)^{D+2\nu +2}$ for non-Neumann boundary conditions with $|\beta
_{j}|\ll a$. The large-distance asymptotic beahvior of the vacuum effective
pressure on the brane at $x^{1}=a_{j}$ is given by (\ref{Pjlarge1}) and (\ref%
{Pjlarge2}) for non-Neumann boundary conditions on that brane and by (\ref%
{PjNlarge}) and (\ref{PjNlarge1}) for Neumann condition. The sign of the
force at large separations depend on the parameter $B_{\nu }$, defined by (%
\ref{Bnu}). For given values of the parameters, the vacuum pressure may also
change its sign as a function of $z$. This means that the forces acting on
different parts of the brane may differ by sign.

A qualitatively different feature of the problem in the AdS bulk is the
presence of the vacuum shear force on the branes. The corresponding part
induced by the second brane is expressed as (\ref{fjint}). It vanishes for
the brane with Dirichlet or Neumann boundary conditions regardless of the
condition on the second brane. Depending on the coefficients in the boundary
conditions, on the separation between the branes and also on the distance
from the AdS boundary, the shear component of the force can be either
positive or negative. At small separations, the leading term in the
expansion of the shear force is given by (\ref{fjsmall2}). In particular,
for a minimally coupled field with negative value of the Robin coefficient $%
\beta _{j}$ the shear force on the brane $x^{1}=a_{j}$ is directed toward
the AdS horizon for non-Dirichlet boundary conditions on the second brane
and toward the AdS boundary for Dirichlet condition. At large separations
the asymptotic for the interaction part of the shear force is given by (\ref%
{fjlarge2}). For minimally and conformally coupled fields and for $\beta
_{j}<0$, at large separations the force $f_{j}^{\mathrm{(int)}}$ is directed
toward the AdS horizon for Neumann boundary condition on the second brane
and toward the AdS boundary for non-Neumann conditions.

\section*{Acknowledgments}

A.A.S. was supported by the grants No. 20RF-059 and No. 21AG-1C047 of the
Science Committee of the Ministry of Education, Science, Culture and Sport
RA. V.Kh.K. was supported by the grant No. 21AG-1C069 of the Science
Committee of the Ministry of Education, Science, Culture and Sport RA.
A.A.S. gratefully acknowledges the hospitality of the INFN, Laboratori
Nazionali di Frascati (Frascati, Italy), where a part of this work was done.

\appendix

\section{Integral representation for the series over eigenvalues}

\label{sec:Ap}

In this section we provide an integral representation for the series $%
S(b,\Delta t,x^{1},x^{\prime 1})$, given by (\ref{S2}). The transformation
will be based on the summation formula \cite{Rome02}
\begin{eqnarray}
\sum_{n=1}^{\infty }\frac{f(u_{n})}{N_{n}} &=&\frac{1}{\pi }\int_{0}^{\infty
}du\,f(u)+\frac{i}{\pi }\int_{0}^{\infty }du\frac{f(e^{\pi i/2}u)-f(e^{-\pi
i/2}u)}{\tilde{c}_{1}(u)\tilde{c}_{2}(u)e^{2u}-1}  \notag \\
&&-\frac{f(0)/2}{1-b_{2}-b_{1}}-\frac{\theta (b_{j})}{2b_{j}}\left[
h_{1}(e^{\pi i/2}/b_{j})+h_{1}(e^{-\pi i/2}/b_{j})\right] ,  \label{APf}
\end{eqnarray}%
where $\theta (x)$ is the Heaviside step function, $\tilde{c}%
_{j}(u)=(b_{j}u-1)/(b_{j}u+1)$ and $h(u)=(b_{j}^{2}u^{2}+1)f(u)$. In (\ref%
{APf}) it is assumed that the function $f(u)$ is analytic in the right
half-plane $\mathrm{Re\,}u>0$. For the series in (\ref{Wf2}) the function $%
f(u)$ is given by the expression
\begin{equation}
f(u)=\frac{e^{-i\sqrt{u^{2}/a^{2}+b^{2}}\Delta t}}{\sqrt{u^{2}/a^{2}+b^{2}}}%
\left[ 2\cos \left( \frac{{u}}{a}{{\Delta x}^{1}}\right) {+}\sum_{l=\pm
1}\left( {e}^{i{|{{x}^{1}{+x}}^{\prime }{^{1}-2{a_{j}}}|u/a}}\frac{iub_{j}-1%
}{iub_{j}+1}\right) ^{l}\right] ,  \label{fu}
\end{equation}%
with $f(0)=0$. By taking into account that for $x>0$ one has%
\begin{equation}
\sqrt{\left( e^{\pm \pi i/2}x\right) ^{2}+b^{2}}=\left\{
\begin{array}{cc}
\sqrt{b^{2}-x^{2}}, & x<b \\
e^{\pm \pi i/2}\sqrt{x^{2}-b^{2}}, & x>b%
\end{array}%
\right. ,  \label{root}
\end{equation}%
and introducing a new integration variable $\lambda =u/a$, the function (\ref%
{S2}) is presented as%
\begin{eqnarray}
S(b,\Delta t,x^{1},x^{\prime 1}) &=&\frac{a}{2}S_{0}(b,\Delta t,{{x}_{-}^{1}}%
)+\frac{a}{4}S_{j}(b,\Delta t,x_{+}^{1})+a\frac{\pi \theta (\beta _{j})}{%
2\beta _{j}}e^{-{|{{x}_{+}^{1}-2{a_{j}}}|/\beta }_{j}}\sum_{l=\pm 1}\frac{%
e^{-i\sqrt{(li/\beta _{j})^{2}+b^{2}}\Delta t}}{\sqrt{(li/\beta
_{j})^{2}+b^{2}}}  \notag \\
&&+\frac{a}{2}\int_{b}^{\infty }d\lambda \frac{2\cosh \left( {\lambda {x}%
_{-}^{1}}\right) {+}\sum_{l=\pm 1}\left[ {e}^{{|{{x}_{+}^{1}-2{a_{j}}}%
|\lambda }}c_{j}(\lambda a)\right] ^{l}}{\left[ c_{1}(\lambda
a)c_{2}(\lambda a)e^{2a\lambda }-1\right] \sqrt{\lambda ^{2}-b^{2}}}\cosh
\left( \sqrt{\lambda ^{2}-b^{2}}\Delta t\right) ,  \label{S3}
\end{eqnarray}%
where $x_{\pm }^{1}=x^{1}\pm x^{\prime 1}$ and%
\begin{eqnarray}
S_{0}(b,\Delta t,{{x}_{-}^{1}}) &=&\int_{0}^{\infty }d\lambda \,\frac{e^{-i%
\sqrt{\lambda ^{2}+b^{2}}\Delta t}}{\sqrt{\lambda ^{2}+b^{2}}}\cos \left( {%
\lambda {x}_{-}^{1}}\right) ,  \notag \\
S_{j}(b,\Delta t,x_{+}^{1}) &=&\int_{0}^{\infty }d\lambda \,\frac{e^{-i\sqrt{%
\lambda ^{2}+b^{2}}\Delta t}}{\sqrt{\lambda ^{2}+b^{2}}}\sum_{l=\pm 1}\left(
{e}^{i{|{{x}_{+}^{1}-2{a_{j}}}|\lambda }}\frac{i\lambda \beta _{j}-1}{%
i\lambda \beta _{j}+1}\right) ^{l}.  \label{S0j}
\end{eqnarray}

For the further transformation of the function $S_{j}(b,\Delta t,x_{+}^{1})$
we rotate the integration contour by the angle $\pi /2$ for the $l=1$ term
and by the angle $-\pi /2$ for the term with $l=-1$. This choice for the
integration contours is dictated by the behavior of the integrands in the
upper and lower half-planes of the complex variable $\lambda $. The poles $%
\lambda =\pm i/\beta _{j}$ for $\beta _{j}>0$ are excluded by semicircles in
the right half-plane $\mathrm{Re\,}\lambda \geq 0$ with small radius. Again,
by using (\ref{root}), this gives%
\begin{eqnarray}
S_{j}(b,\Delta t,x_{+}^{1})&=&2\int_{b}^{\infty }d\lambda \frac{\cosh \left(
\sqrt{\lambda ^{2}-b^{2}}\Delta t\right) }{\sqrt{\lambda ^{2}-b^{2}}}\frac{%
e^{-{\lambda |{{x}_{+}^{1}-2{a_{j}}}|}}}{c_{j}(\lambda a)}  \notag \\
&& -\frac{2\pi }{\beta _{j}}\theta (\beta _{j})\sum_{l=\pm 1}\frac{e^{-i%
\sqrt{\left( i/\beta _{j}\right) ^{2}+b^{2}}\Delta t}}{\sqrt{\left( i/\beta
_{j}\right) ^{2}+b^{2}}}e^{-{|{{x}_{+}^{1}-2{a_{j}}}|/}\beta _{j}}.
\label{Sj}
\end{eqnarray}%
Substituting this in (\ref{S3}) we see that the terms with the Heaviside
step function are cancelled out and the function (\ref{S2}) is expressed as%
\begin{eqnarray}
&& S(b,\Delta t,x^{1},x^{\prime 1}) =\frac{a}{2}S_{0}(b,\Delta t,{{\Delta x}%
^{1}})+\frac{a}{2}\int_{b}^{\infty }d\lambda \frac{\cosh \left( \sqrt{%
\lambda ^{2}-b^{2}}\Delta t\right) }{\sqrt{\lambda ^{2}-b^{2}}}\frac{e^{-{%
\lambda |{{x}_{+}^{1}-2{a_{j}}}|}}}{c_{j}(\lambda a)}  \notag \\
&& \qquad +\frac{a}{2}\int_{b}^{\infty }d\lambda \frac{2\cosh \left( {%
\lambda {x}_{-}^{1}}\right) {+}\sum_{l=\pm 1}\left[ {e}^{{|{{x}_{+}^{1}-2{%
a_{j}}}|\lambda }}c_{j}(\lambda a)\right] ^{l}}{\left[ c_{1}(\lambda
a)c_{2}(\lambda a)e^{2a\lambda }-1\right] \sqrt{\lambda ^{2}-b^{2}}}\cosh
\left( \sqrt{\lambda ^{2}-b^{2}}\Delta t\right) .  \label{S4}
\end{eqnarray}%
Another representation, symmetric with respect to the branes, is obtained
from (\ref{S4}) combining the integrals:%
\begin{eqnarray}
S(b,\Delta t,x^{1},x^{\prime 1}) &=&\frac{a}{2}S_{0}(b,\Delta t,{{\Delta x}%
^{1}})+\frac{a}{2}\int_{b}^{\infty }d\lambda \frac{\cosh \left( \sqrt{%
\lambda ^{2}-b^{2}}\Delta t\right) }{\sqrt{\lambda ^{2}-b^{2}}}  \notag \\
&&\times \frac{2\cosh \left( {\lambda {x}_{-}^{1}}\right) {+}\sum_{j=1,2}{e}%
^{{|{{x}_{+}^{1}-2{a_{j}}}|\lambda }}c_{j}(\lambda a)}{c_{1}(\lambda
a)c_{2}(\lambda a)e^{2a\lambda }-1}.  \label{S5}
\end{eqnarray}%
Note that in the limit $(-1)^{j^{\prime }}a_{j^{\prime }}\rightarrow +\infty
$, with $j^{\prime }=1$ for $j=2$ and $j^{\prime }=2$ for $j=1$, the last
term in (\ref{S4}) goes to zero and the first two terms in the right-hand
side determine the contribution to the Wightman function in the geometry of
a single bran at $x^{1}=a_{j}$.

\end{document}